\documentclass[fleqn,usenatbib]{mnras}

\usepackage{newtxtext,newtxmath}
\usepackage{nicematrix}
\usepackage{booktabs}
\usepackage{tikz}
\usepackage{multirow}
\usepackage{fontawesome5}
\usepackage[outline]{contour}

\usepackage{tabularx}
\newcolumntype{Y}{>{\centering\arraybackslash}X} 
\newcolumntype{R}{>{\raggedleft\arraybackslash}X} 
\usepackage{multirow, booktabs, graphicx}
\usepackage{colortbl}
\usepackage[table,xcdraw]{xcolor}

\usepackage[T1]{fontenc}

\DeclareRobustCommand{\VAN}[3]{#2}
\let\VANthebibliography\thebibliography
\def\thebibliography{\DeclareRobustCommand{\VAN}[3]{##3}\VANthebibliography}

\usepackage{graphicx}	
\usepackage{amsmath}	

\title[Illuminating the Invisible]{The Challenge in Illuminating the Invisible: Constraining LyC Escape with Bayesian Modelling and Symbolic Regression}

\author[A. Stoffers et al]{Amanda Stoffers,$^{1, 2}$\thanks{E-mail: aas208@cam.ac.uk}
Sandro Tacchella,$^{1,2}$, Charlotte Simmonds$^{1,2,3}$, Benjamin D. Johnson,$^{4}$ 
\newauthor{Roberto Maiolino$^{1,2,5}$}
\\\\
$^{1}$Kavli Institute for Cosmology, University of Cambridge, Madingley Road, Cambridge CB3 0HA, UK\\
$^{2}$Cavendish Laboratory, University of Cambridge, 19 JJ Thomson Avenue, Cambridge CB3 0HE, UK\\
$^{3}$Departamento de Astronomía, Universidad de Chile, Camino El Observatorio 1515, Las Condes, Santiago, Chile\\
$^{4}$Center for Astrophysics | Harvard \& Smithsonian, 60 Garden Street, Cambridge, MA 02138, USA\\
$^{5}$ Department of Physics and Astronomy, University College London, Gower Street, London WC1E 6BT, UK
}

\date{Accepted XXX. Received YYY; in original form ZZZ}

\pubyear{\the\year{}}

\begin{document}
\label{firstpage}
\pagerange{\pageref{firstpage}--\pageref{lastpage}}
\maketitle

\begin{abstract}
Direct observations of Lyman continuum (LyC) radiation from galaxies during the Epoch of Reionization (EoR) are impeded by absorption in the intergalactic medium, requiring indirect methods to infer the escape fraction of ionizing photons ($f_\mathrm{esc}^{\mathrm{LyC}}$). One approach is to develop and validate such methods on local analogues of the high-redshift galaxies with directly detected LyC leakage. In this study, we constrain $f_\mathrm{esc}^\mathrm{LyC}$ using a Bayesian spectral energy distribution (SED) fitting framework built on \texttt{Prospector}, which incorporates a non-parametric star-formation history, a flexible dust attenuation curve, self-consistent nebular emission, and fiber aperture-loss corrections. Our methodology jointly fits broadband photometry and emission line fluxes. We apply six models to the Low-redshift LyC Survey (LzLCS), a sample of local galaxies with physical properties comparable to EoR galaxies, and evaluate them based on their ability to recover the observed flux and their relative Bayesian evidence. The best-performing model is further assessed through a parameter recovery test, demonstrating that $f_\mathrm{esc}^\mathrm{LyC}$ can be recovered within uncertainties. Building on these results, we present updated $f_\mathrm{esc}^\mathrm{LyC}$ estimates for the LzLCS sample, with a median of 4\%, and values reaching as high as 51\%, with 26 of 64 galaxies having a cosmologically relevant $f_\mathrm{esc}^\mathrm{LyC}$ ($\gtrsim5\%$). Additionally, we present a revised UV $\beta$-slope vs $\log_{10}(f_\mathrm{esc}^\mathrm{LyC})$ relation, derived using symbolic regression with \texttt{PySR} trained on a synthetic dataset generated with \texttt{Prospector}:$\log_{10}(f_\mathrm{esc}^\mathrm{LyC}) = (-2.30 \pm 1.28)\beta - (6.26 \pm 2.91)$, ($\sigma = 0.43$~dex). The relation successfully reproduces the $f_\mathrm{esc}^\mathrm{LyC}$ obtained from full SED fitting of the LzLCS sample within $1\sigma$.

\end{abstract}

\begin{keywords}
dark ages, reionization, first stars -- galaxies: star formation -- galaxies: stellar content -- galaxies: evolution -- ultraviolet: galaxies
\end{keywords}


\begingroup
\let\clearpage\relax
\endgroup
\newpage

\section{Introduction}
\label{sec:introduction}

The Epoch of Reionization (EoR) marks the Universe’s last major hydrogen phase transition, during which the intergalactic medium (IGM) was transformed from neutral to ionized. Observations constrain the end of this process to $z \approx 5.7$ \citep{bosman_hydrogen_2022, keating_long_2020}, but its exact time evolution and topology remains uncertain. The progression of reionization depends on three key quantities: the number density of ionizing sources, their efficiency in producing hydrogen-ionizing photons with $E > 13.6~\mathrm{eV}$ ($\xi_\mathrm{ion}$), and the fraction of these photons that escape ($f_\mathrm{esc}^\mathrm{LyC}$) out of the system into the IGM. Because these quantities depend on galaxy mass, metallicity, and redshift, different assumptions lead to very different reionization histories. For instance, should reionization be driven by numerous faint sources ($M_\mathrm{UV} > -17$ AB mag) with low $f_\mathrm{esc}^\mathrm{LyC}$, it would have begun earlier and proceeded gradually. Conversely, if dominated by rarer, luminous sources with high $f_\mathrm{esc}^\mathrm{LyC}$, reionization would have started later and progressed more rapidly. Constraining the population distributions of these parameters is therefore important for understanding the evolution of the EoR. 

The production rate of ionizing photons of individual galaxies can be estimated with stellar population synthesis models \citep{kennicutt_global_1998, simmonds_ionizing_2023, simmonds_impact_2024,  pahl_spectroscopic_2025, Papovich:2025aa} or inferred directly from observations, most robustly via spectroscopic measurements \citep{Schaerer:2016aa, tang_mmtmmirs_2019}, while recent \textit{James Webb Space Telescope} (JWST) results provide measurements of the number density of sources at $z>9$ \citep{harikane_jwst_2025, whitler_z_2025}. In contrast, constraining $f_\mathrm{esc}^\mathrm{LyC}$ remains far more difficult, owing to the complex and inhomogeneous structure of the inter stellar medium (ISM) \citep{gazagnes_neutral_2018}. In addition to the complex ISM structures that hinder estimations of $f_\mathrm{esc}^\mathrm{LyC}$, the intervening neutral hydrogen makes LyC detection essentially impossible at $z>4$ \citep{inoue_updated_2014}. As a result, studies must rely on indirect tracers. 

Common diagnostics include the UV continuum slope ($\beta$, defined as $f_\lambda \propto \lambda^\beta$; \citealt{calzetti_dust_1994}) in combination with the H$\beta$ equivalent width \citep{Zackrisson:2013aa, Zackrisson:2017aa} and emission line ratios such as [O\,\textsc{iii}]/[O\,\textsc{ii}] ($O_{32}$) \citep{izotov_low-redshift_2018, Mascia:2023aa}. Additional indicators are Ly$\alpha$ peak separation \citep{verhamme_using_2015, izotov_diverse_2020}, Ly$\alpha$ equivalent width \citep{pahl_redshift_2020, gazagnes_origin_2020, saldana-lopez_low-redshift_2022, flury_low-redshift_2022}, and Mg\,\textsc{ii} doublet ratios \citep{chisholm_optically_2020, katz_mg_2022}. Other proposed tracers include strong C\,\textsc{iv} emission \citep{schaerer_revealing_2022, saxena_strong_2022}, [S\,\textsc{ii}] deficiency \citep{Borthakur:2014aa, Alexandroff:2015aa, wang_new_2019, wang_low-redshift_2021}, low-ionization absorption line (LIS) kinematics \citep{Jaskot:2017aa}, Si II and Si III LIS outflow diagnostics \citep{chisholm_galaxies_2017}, HI-covering fraction estimations via metal absorption lines and UV Hi \citep{gazagnes_neutral_2018, gazagnes_origin_2020}. While theoretical studies predict that connecting metallic absorption lines to $f_\mathrm{esc}^\mathrm{LyC}$ mainly shows promise in extreme cases (e.g., $f_\mathrm{esc}^\mathrm{LyC} = 0$ or $>30\%$) \citep{mauerhofer_uv_2021}, \cite{saldana-lopez_low-redshift_2022} find that combined measurements of UV absorption lines with dust attenuation are applicable over a wide range of lower $f_\mathrm{esc}^\mathrm{LyC}$ (0.02\% - 20\%). Furthermore, the large scatter in single–parameter predictors of $f_\mathrm{esc}^\mathrm{LyC}$ as demonstrated by the highly dispersed trends \citep[e.g.][]{flury_low-redshift_2022} motivates a transition toward multivariate approaches. The first such methods were introduced by \citet{Mascia:2023aa, Mascia:2024aa}, while techniques that account for censoring include survival-analysis frameworks \citep{Jaskot_2024}. Spectral energy distribution (SED) fitting provides a complementary, self-consistent approach by incorporating multi-wavelength photometry and spectroscopy \citep{Giovinazzo:2025aa}. A comprehensive overview of current methods is given in 
\citet{jaskot_ionizing_2025}.

To calibrate and test such tracers, local analogues of EoR galaxies are widely used as calibration benchmarks. Compact ($r_{50,~\mathrm{NUV}} \leq 1~\mathrm{kpc}$), star-forming ($\Sigma_\mathrm{SFR}\geq 0.1 M_\odot~\mathrm{yr}^{-1}\mathrm{kpc}^{-2}$), low-metallicity systems are the most widely used candidates \citep{flury_low-redshift_2022}, though their comparability to high-redshift populations remains debated. Discrepancies between high- and low-redshift estimations $f_\mathrm{esc}^\mathrm{LyC}$ may arise from either a redshift evolution in the underlying relations or from differences in the galaxy populations probed at different epochs \citep{saldana-lopez_vandels_2023, 2024ApJ...974..212P}. For example, \citet{citro_challenging_2025} report non-detections of LyC in galaxies at $z \approx 2.3$ that exhibit strong Ly$\alpha$ emission and very blue UV continuum slopes ($\beta \sim -2.4$). Nevertheless, local analogues can provide valuable laboratories for probing the mechanisms that regulate LyC escape.
Spatially resolved observations of local galaxies, such as \textit{Haro 11} at $z=0.02$ \citep{komarova_haro_2024}, reveal that the regions dominating the observed LyC flux are not necessarily the same as the regions driving the emission of indirect tracers like [O\,\textsc{iii}] or H$\alpha$ \citep{ostlin_source_2021}. This implies that the spatial origin of leaking LyC does not necessarily coincide with the regions dominating nebular tracers such as [O III] or H$\alpha$, offering a natural explanation for the large scatter observed in correlations between $f_\mathrm{esc}^\mathrm{LyC}$ and indirect indicators.
It is important to note that even for low-redshift galaxies, there is no such thing as a directly observed $f_\mathrm{esc}^\mathrm{LyC}$. Since $f_\mathrm{esc}^\mathrm{LyC}$ is defined as the ratio of escaped to intrinsically produced LyC photons, assumptions about the intrinsic production are always required. This introduces strong model dependencies and makes it difficult to compare $f_\mathrm{esc}^\mathrm{LyC}$ across different analyses. Moreover, line-of-sight effects can further complicate the picture, as leakage along one direction may not reflect the global escape fraction of the galaxy.

An alternative approach to study $f_\mathrm{esc}^\mathrm{LyC}$ is through numerical simulations, but translating their results into observationally testable predictions is challenging. Numerous radiation hydrodynamics simulations show that the escape of H-ionizing photons highly depends on the properties of the host galaxy and is highly anisotropic \citep{paardekooper_first_2015, barrow_lyman_2020, rosdahl_lyc_2022, smith_physics_2022}.  Line-of-sight $f_\mathrm{esc}^\mathrm{LyC}$ are not Gaussian-distributed around the global escape fraction. Instead, galaxies often appear non-leaking along most lines of sight, even if their integrated escape fraction is nonzero \citep{mauerhofer_uv_2021}.
The complex, time-dependent relation of $f_\mathrm{esc}^\mathrm{LyC}$ seems to be closely connected to the presence of both radiative and mechanical stellar feedback \citep{kimm_escape_2014, trebitsch_fluctuating_2017}. 
An additional challenge of approaching $f_\mathrm{esc}^\mathrm{LyC}$ with simulations lies in the range of resolution that is necessary to simulate star-forming clouds in detail.
Also, determining $f_\mathrm{esc}^\mathrm{LyC}$ through forward modelling of galaxy structures and ionizing photon transport with numerical simulations is highly dependent on assumptions about galaxy formation, feedback prescriptions, the treatment of the multiphase ISM, and dust attenuation. Predictions from such simulations generally indicate that a global average escape fraction of $5$–$20\%$ is required to reproduce the timeline of cosmic reionization \citep{trebitsch_reionization_2022}. When obscured star formation is accounted for, the predicted escape fractions are lower, with values of $5$–$10\%$ found in \textsc{Thesan} \citep{yeh_thesan_2023} and \textsc{Obelisk} \citep{trebitsch_obelisk_2021}, and below $5\%$ in \textsc{Sphinx} \citep{rosdahl_lyc_2022}. The \textsc{Sphinx} simulations also show that escape fractions fluctuate strongly within individual galaxies on Myr timescales, regulated by the interplay of supernovae and radiative feedback. Beyond global escape fractions, radiative transfer simulations of the ISM have highlighted that line ratios, diffuse ionized gas, and related diagnostics provide powerful probes of ISM physics and feedback processes, and help identify which stellar populations dominate the ionization budget \citep{tacchella_h_2022, smith_physics_2022, mcclymont_nature_2024, Choustikov:2024aa}.  

In this paper, we use SED modelling to re-derive the stellar population properties of the Low-z LyC Survey \citep{flury_low-redshift_2022-1} galaxies, with a particular focus on their $f_\mathrm{esc}^\mathrm{LyC}$. We (a) build a physically flexible SED model that explicitly includes birth-cloud vs diffuse dust and a parameter for runaway ionizing photons, (b) test multiple prior and dust prescriptions, and (c) assess their reliability with parameter recovery tests. 

A concise overview of the dataset is presented in Section~\ref{sec:data}. In Section~\ref{sec:models}, we describe the set of models adopted for fitting each galaxy and quantify the impact of different priors and model assumptions. Section~\ref{sec:results} presents our re-derived estimates of $f_\mathrm{esc}^\mathrm{LyC}$ and other galaxy properties, which we compare to recent high-redshift observations from JWST to assess whether the LzLCS galaxies provide suitable local analogues of EoR systems. In Section~\ref{sec:symreg}, we revisit correlations between $f_\mathrm{esc}^\mathrm{LyC}$ and the newly inferred stellar properties by employing symbolic regression to derive an analytic relation linking $f_\mathrm{esc}^\mathrm{LyC}$ to directly observable quantities. Finally, we discuss the challenges of our analysis and future work in Section~\ref{sec:discussion} and summarize our findings in Section~\ref{sec:conclusion}. 

Throughout this work, we assume a \textit{Planck} 2018 flat $\Lambda$CDM cosmology with $H_0 = 67.36\,\mathrm{km\,s^{-1}\,Mpc^{-1}}$, $\Omega_m = 0.3153$, and $\Omega_\Lambda = 0.6847$ \citep{planck_collaboration_planck_2020}. Reported parameter values correspond to the posterior median and 16th–84th percentile range. For derived quantities (i.e., quantities computed from fitted parameters), we evaluate them across the full posterior distribution and quote the median and percentiles of the resulting distribution.


\section{Data}
\label{sec:data}

The Cosmic Origins Spectrograph (COS) on the Hubble Space Telescope (HST) enables the detection of escaping LyC photons from galaxies at  $z \sim 0.3$ , leading to the assembly of the Low-Redshift Lyman Continuum Survey (LzLCS; PI: Jaskot, HST Project ID: 15626, \citealt{flury_low-redshift_2022-1}). The galaxies in this sample meet several criteria that suggest they are suitable analogs for high-redshift Reionization-era galaxies: they are metal-poor ($12 + \log_{10}(O/H) \sim 8.1$), actively star-forming and compact ($\Sigma_\mathrm{SFR} > 0.1~M_\odot \mathrm{yr}^{-1} \mathrm{kpc}^{-2}$). We discuss how their properties are in line with high-redshift galaxies in Section~\ref{sec:results}. In this work, we reanalyse the LzLCS using a more flexible modelling framework than the original study \citep{flury_low-redshift_2022-1}, with the aim of deriving a consistent set of galaxy properties. In contrast to the initial analysis, our Bayesian fitting approach simultaneously incorporates both photometry and emission-line measurements, allowing for a more comprehensive and self-consistent characterization of each galaxy.
\subsection{Sample Definition and Selection Criteria}
\label{subsec:sample}

The galaxy sample was selected from optical spectroscopy in the 16th Data Release of the Sloan Digital Sky Survey (SDSS; \citealt{2020ApJS..249....3A}) and far-ultraviolet (FUV) photometry from GALEX \citep{Martin_2005}. The redshift range was chosen such that the Lyman continuum (LyC) falls within the more sensitive region of the \textit{HST}/COS detector. Details of the data reduction are provided in \citet{flury_low-redshift_2022-1}. While the SDSS and GALEX photometry are aperture-matched, we acknowledge that the use of unmatched SDSS fiber spectroscopy introduces potential aperture mismatches due to differing spatial coverage. We account for these fiber / aperture mismatches in the modelling via a free photometry–spectroscopy scaling parameter ($f_\mathrm{scale}$; Section~\ref{sec:models}).

The LzLCS was explicitly designed to test three proposed diagnostics of LyC escape and combinations thereof:  
\begin{itemize}
    \item \textbf{Star formation rate surface density:} $\Sigma_\mathrm{SFR} > 0.1\,M_\odot\,\mathrm{yr}^{-1}\,\mathrm{kpc}^{-2}$, characteristic of compact starbursts where feedback may facilitate LyC leakage \citep{heckman_escape_2001, clarke_galactic_2002}.  
    \item \textbf{Ionization state:} $O32 = [{\rm O\,III}]\,\lambda5007 / [{\rm O\,II}]\,\lambda3727 \geq 3$, indicative of hard ionizing spectra and proposed as an indirect tracer of $f_\mathrm{esc}^\mathrm{LyC}$ \citep{Jaskot:2013aa, izotov_low-redshift_2018}.  
    \item \textbf{UV continuum slope:} $\beta < -2$, derived from GALEX FUV--NUV colours, consistent with young, dust-poor stellar populations \citep{wilkins_accuracy_2012, wilkins_interpreting_2013, calabro_vandels_2021}.  
\end{itemize}
While these criteria target galaxies with conditions thought to favor LyC escape, they also introduce a selection bias, meaning that any correlations derived between $f_\mathrm{esc}^\mathrm{LyC}$ and galaxy properties may only reflect this pre-selected, extreme subset of systems (e.g.\ high-O32 galaxies) if the selection is not specifically accounted for. Moreover, galaxies with similar $\Sigma_\mathrm{SFR}$, $O32$, and $\beta$ can still exhibit a wide range of $f_\mathrm{esc}^\mathrm{LyC}$ values \citep{izotov_low-redshift_2018, flury_low-redshift_2022-1}, highlighting the limitations of relying on these diagnostics alone.

\subsection{Photometry and Emission Lines}
\label{subsec:photEL}
The uncorrected photometric and emission line data were obtained from \citet{flury_low-redshift_2022-1}. We applied corrections for foreground Milky Way reddening to both datasets using the dust maps from \citet{m_green_dustmaps_2018} and the Milky Way extinction curve of \citet{fitzpatrick_correcting_1999}, following the methodology adopted in the original survey.
The photometric data comprise observations in the SDSS \textit{u, g, r, i, z} bands and the GALEX NUV and FUV filters. The data quality is generally high, with signal-to-noise ratios (SNR) ranging from 91 in the FUV to over 700 in the \textit{r}-band.
We include the following emission lines fluxes and their measured uncertainties in our analysis:[O\,\textsc{ii}]~$\lambda\lambda3726,3729$, [Ne\,\textsc{iii}]~$\lambda3869$, He\,\textsc{i}~$\lambda3889$, [Ne\,\textsc{iii}]~$\lambda3968$, H$\epsilon$~$\lambda3970$, H$\delta$~$\lambda4102$, H$\gamma$~$\lambda4341$, [O\,\textsc{iii}]~$\lambda4363$, H$\beta$~$\lambda4861$, [O\,\textsc{iii}]~$\lambda4959$, [O\,\textsc{iii}]~$\lambda5007$, He\,\textsc{i}~$\lambda5876$, [O\,\textsc{i}]~$\lambda6300$, [N\,\textsc{ii}]~$\lambda6548$, H$\alpha$~$\lambda6563$, [N\,\textsc{ii}]~$\lambda6584$, and [S\,\textsc{ii}]~$\lambda\lambda6716,6731$. 
[O\,\textsc{i}]~$\lambda6300$ and [O\,\textsc{iii}]~$\lambda4363$ are very faint for most galaxies and therefore excluded from the fit. Notably, the key diagnostic lines for ionization state modeling -- [O\,\textsc{iii}]~$\lambda4959$, [O\,\textsc{iii}]~$\lambda5007$, and H$\beta$~$\lambda4861$ -- exhibit excellent median SNRs of approximately 65, 76, and 50, respectively. For the analysis that follows, we cap the SNR at 20 for both photometry and emission-line fluxes, acknowledging that our models are not yet able to fully exploit higher-SNR data in a physically reliable way. In addition, a small fraction ($<10\%$) of the LzLCS galaxies may be affected by SDSS emission-line clipping \citep{flury_low-redshift_2022-1}; we mitigate this by adopting an error floor on nebular line fluxes, with minimal possible impact on the inferred parameters given that the SED fits are constrained by the full set of emission lines and photometric data.

While some galaxies in the sample have additional HST photometry that could further tighten constraints on the inferred stellar properties, our aim is to develop and validate a framework applicable to high-redshift datasets, where comparable high-resolution UV imaging is typically unavailable. We therefore restrict our analysis to the uniformly available SDSS and GALEX photometry. Incorporating HST and acquisition-image data for the relevant subsample would be a natural extension of this work.

\subsection{Measuring the Lyman Continuum}
\label{subsubsec:measuringTheLymanConituum}

In the LzLCS, \citet{flury_low-redshift_2022} excluded all wavelengths longward of 1180\,\AA\ (observed frame) to remove contamination from geocoronal Ly$\alpha$ and N\,\textsc{i} $\lambda$1200. The LyC flux was then measured in a 20\,\AA\ window positioned as close as possible to $\lambda_\mathrm{rest}$~=~900\AA. The interval between $\lambda_\mathrm{rest}$~=~900–912\,\AA\ was further excluded because of scattered starlight in the HST/COS optics. As a result, the LyC flux is typically measured between $\lambda_\mathrm{rest}$~=~840 and $\lambda_\mathrm{rest}$~=~890, depending on the galaxy. Throughout this paper, the SED-inferred LyC fluxes are evaluated over the same wavelength range when they are compared to observational measurements.
\citet{flury_low-redshift_2022-1} estimated the escape fraction from the UV SED ($f_\mathrm{esc, ~UV}^\mathrm{LyC}$), the H$\beta$ EW ($f_{\mathrm{esc,~H}\beta}^\mathrm{LyC}$) and the ratio between ionizing and non-ionizing UV flux $F_{900}^\mathrm{rest}/F_{1100}^\mathrm{rest}$. While this flux ratio can serve as a proxy for f$_\mathrm{esc}^\mathrm{LyC}$ \citep{wang_new_2019}, other studies have argued that this wavelength range is unsuitable for estimating $f_\mathrm{esc,~UV}^\mathrm{LyC}$ due to potential nebular LyC near $\lambda \sim 900\,\text{\AA} $ \citep{simmonds_impact_2024, izotov_great_2025}.


\section{Inferring the Escape Fraction Indirectly}
\label{sec:models}

To infer $f_\mathrm{esc}^\mathrm{LyC}$ without access to the ionizing spectrum, we model the LzLCS data with the Bayesian SED fitting code \texttt{Prospector} \citep{johnson_stellar_2021}, which allows for a simultaneous and self-consistent treatment of photometry and emission lines. In this section, we outline the theoretical framework and modelling choices, describe the implementation of $f_\mathrm{esc}^\mathrm{LyC}$, and specify the adopted priors. We then investigate the impact of varying dust prescriptions, apply different priors on the runaway fraction of young, hot (10,000 K to 50,000 K) OB stars ($f_\mathrm{run}^\mathrm{OB}$), and test the inclusion of a parameter accounting for fiber-loss corrections. The different models are summarized in Tab.~\ref{tab:models}. A parameter recovery test validating the framework is provided in Appendix~\ref{sec:parameterRecoveryTest}.\\

\begin{figure}
    \centering
    \includegraphics[width=\linewidth]{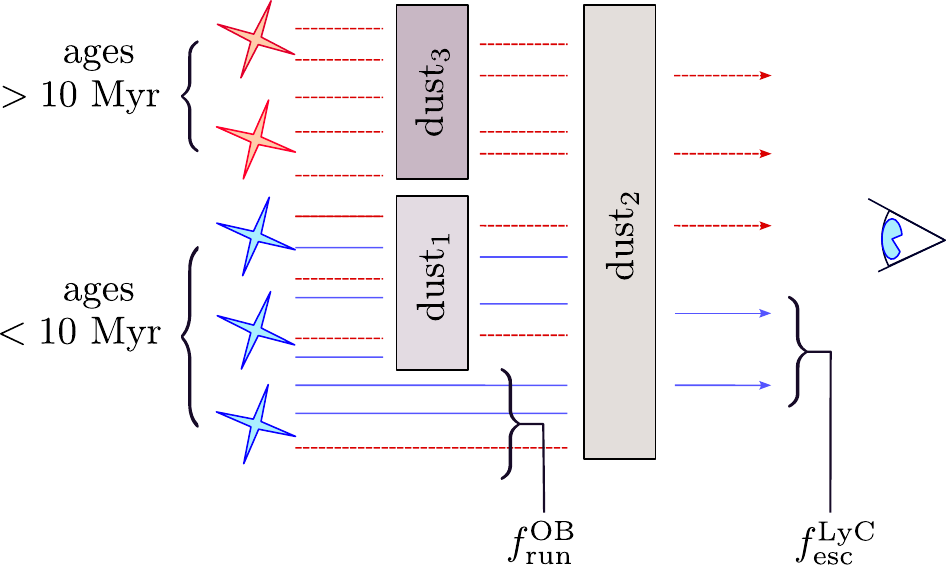}
    \caption{Schematic illustration of the three dust components in our models. Red and blue stars represent stellar populations older and younger than $10$ Myr, respectively. Red dashed lines indicate non-ionizing radiation, while blue lines indicate ionizing radiation. Light from the older stellar population is attenuated by dust$_3$ (optical depth $\tau_3$). For the young population, a fraction of the ionizing photons is absorbed by dust$_1$, while the fraction escaping this birth-cloud dust and nebular absorption is described by $f_\mathrm{run}^\mathrm{OB}$. All stellar light then passes through the diffuse dust screen dust$_2$. The fraction of ionizing photons that survives all dust attenuation corresponds to $f_\mathrm{esc}^\mathrm{LyC}$.}
    \label{fig:dustcomponents}
\end{figure}

Each model is described by 15 or more free parameters. We directly fit for the logarithm of the stellar mass $\log_{10}(M_\star/M_\odot)$ (\texttt{logmass}), assuming a uniform prior over the range $[6, 12]$, and the logarithm of the stellar metallicity $\log_{10}(Z/Z_\odot)$ (\texttt{logzsol}), for which we adopt a clipped normal prior between $-2.0$ and $0.19$, centered at $\mu = -1.0$ with a standard deviation of $\sigma = 0.3$. We assume a \citet{Chabrier:2003aa} initial mass function (IMF).

We adopt a two-component dust attenuation model following \citet{charlot_simple_2000}. The first component describes attenuation of young stars ($<$10 Myr) and their associated nebular emission within dusty birth clouds, modelled as a power-law with fixed slope $-1$. 
The second component represents a diffuse dust screen that attenuates all stellar and nebular emission across the galaxy. Its attenuation curve follows the prescription of \citet{kriek_dust_2013}, where the slope of the attenuation law is tied to the UV bump strength. The parameter \texttt{dust\_index} is modelled as an offset relative to the  \citet{calzetti_dust_1994} attenuation curve, with a uniform prior over $[-1, 4]$. For the diffuse optical depth ($\tau_2$), we adopt a clipped normal prior between $0.0$ and $4.0$ with $\mu = 0.3$ and $\sigma = 1.0$, consistent with observational constraints \citep{calzetti_dust_1994, price_direct_2014}. When fitting for the birth cloud dust component in this dust setup (\texttt{d2}), we fit for the ratio of optical depths between the young and diffuse dust screens ($\tau_1/\tau_2$), assuming a clipped normal prior centered at 1.0.

In a subset of models (\texttt{d3}), we extend this to a three-component dust prescription (Fig.~\ref{fig:dustcomponents}): (i) a dust screen around older stars ($>10$ Myr) with optical depth $\tau_3$, (ii) a young stellar birth-cloud component defined via $\tau_1/\tau_3$, and (iii) a diffuse dust screen defined via $\tau_2/\tau_3$. For the older stellar population, we adopt a clipped normal prior with $\mu = 0.3$ and $\sigma = 1.0$, bounded within $[0.0, 4.0]$. For the young stellar component ($\tau_1/\tau_3$), we use a clipped normal prior centered at 1.0 with $\sigma = 0.3$, restricted to $[0.0, 2.0]$. Finally, for the diffuse screen ($\tau_2/\tau_3$), we assume a clipped normal prior with $\mu = 0.4$ and $\sigma = 0.3$, also bounded within $[0.0, 2.0]$, reflecting expectations of weaker attenuation from the diffuse ISM compared to local birth-cloud dust.

\texttt{Prospector} models both nebular continuum and line emission using the interpolation grids of \citet{byler_nebular_2017}, which are based on \textsc{Cloudy} photoionization models \citep[v13.03;][]{ferland_2013_2013}. The grids span a parameter space defined by: (i) the ionization parameter (\texttt{gas\_logu}), for which we adopt a uniform prior over $[-4, -1]$; (ii) the gas-phase metallicity (\texttt{gas\_logz}), with a uniform prior over $[-2, 0.5]$; and (iii) the age of the stellar population, using stellar spectra of matched age and metallicity as the ionizing source. Nebular emission is scaled to the ionizing photon output predicted at each single stellar population (SSP) age for the specified star formation history (SFH). The grid is truncated at $\log U = -1$, limiting its ability to reproduce the most extreme [O\,\textsc{iii}] ratios \citep[e.g.,][]{ferland_2013_2013, katz_first_2023}. The Cloudy grid adopted in this work neglects dust as an explicit opacity and heating source within the ionized gas, while still accounting for elemental depletion onto dust grains; this simplifying assumption represents a caveat of our analysis and may introduce systematic uncertainty in the inferred LyC escape fractions.

The SFH is modelled non-parametrically across eight lookback-time bins. The first two are fixed at 5 and 10 Myr, while the remaining six are logarithmically spaced back to $z=20$, following \citet{ocvirk_stecmap_2006}, who showed that population separability improves with log-time spacing. Star formation in each bin is parameterized by $\log(\text{SFR}_n/\text{SFR}_{n+1})$, i.e.\ the logarithmic ratio of adjacent bins. We adopt Student-$t$ priors for these ratios, with mean vector $\vec{0}$, scale vector $\vec{0.3}$, and $\nu=2$ degrees of freedom \citep{leja_how_2019}.  

We also include $f_\mathrm{run}^\mathrm{OB}$, the fraction of starlight from young stars that escapes the local birth-cloud attenuation ($\tau_1$) without producing nebular emission, while still being subject to the diffuse dust component.LyC escape is modeled as the fraction of sightlines that avoid nebular absorption but can still experience attenuation by diffuse dust. In dust-free conditions, $f_\mathrm{run}^\mathrm{OB}$ directly traces $f_\mathrm{esc}^\mathrm{LyC}$. In the following subsections, we explore three distinct priors for $f_\mathrm{run}^\mathrm{OB}$. A schematic illustration showing how dust, $f_\mathrm{esc}^\mathrm{LyC}$, and $f_\mathrm{run}^\mathrm{OB}$ are implemented in our model is shown in Fig.~\ref{fig:dustcomponents}.
Lastly, we fit for the parameter $f_\mathrm{scale}$, a free scaling factor between the photometry and line fluxes that accounts for fiber/aperture losses. Losses arise because broadband photometry typically captures the total galaxy light, whereas spectroscopy and emission-line measurements can miss flux depending on the galaxy’s position and extent relative to the spectrograph slit or the fiber. To correct for this mismatch, our fiducial models include $f_\mathrm{scale}$ as a scaling factor between photometric and spectroscopic fluxes. A complete overview of all model parameters and priors is provided in Table~\ref{tab:priors}.

\begin{table*}
\centering
\caption{Summary of the prior choices for the model parameters. Rows shaded in light gray correspond to parameters used in the two-component dust model, while rows shaded in dark gray indicate parameters specific to the three-component dust model. For $f_\mathrm{run}^\mathrm{OB}$ and $f_\mathrm{scale}$, two different priors are tested across different model configurations which are summarised in Tab.~\ref{tab:models}}
\begin{NiceTabular}{lll}
\toprule\toprule
\textbf{Parameter} & \textbf{Description} & \textbf{Prior} \\
\midrule\midrule
\texttt{logmass}       & $\log_{10}(M_{\star}/M_\odot)$                 & Uniform $[6, 12]$ \\
\texttt{logzsol}       & $\log_{10}(Z_{\star}/Z_\odot)$                 & Clipped Normal $[-2.0, 0.19]$, $\mu=-1.0$, $\sigma=0.3$ \\
\texttt{dust\_index}   & Attenuation curve slope of dust 2               & Uniform $[-1, 4]$ \\

\rowcolor{gray!10}
$\tau_2$              & Diffuse dust optical depth             & Clipped Normal $[0.0, 4.0]$, $\mu=0.3$, $\sigma=1.0$ \\
\rowcolor{gray!10}
$\tau_1/\tau_2$       & Young-to-diffuse dust ratio            & Clipped Normal $[0.0, 2.0]$, $\mu=1.0$, $\sigma=0.3$ \\

\rowcolor{gray!20}
$\tau_3$              & Optical depth of the old stars               & Clipped Normal $[0.0, 4.0]$, $\mu=0.3$, $\sigma=1.0$ \\
\rowcolor{gray!20}
$\tau_1/\tau_3$       & Young-to-old dust ratio              & Clipped Normal $[0.0, 2.0]$, $\mu=1.0$, $\sigma=0.3$ \\
\rowcolor{gray!20}
$\tau_2/\tau_3$       & Diffuse-to-old dust ratio            & Clipped Normal $[0.0, 2.0]$, $\mu=0.4$, $\sigma=0.3$ \\

\texttt{gas\_logu}     & Ionization parameter                   & Uniform $[-4, -1]$ \\
\texttt{gas\_logz}     & Gas-phase metallicity                  & Uniform $[-2, 0.5]$ \\
$\log(\texttt{SFR}_n/\texttt{SFR}_{n+1})$ & SFR bin ratios     & Student-t, $\mu=0$, scale $0.3$, dof $=2$ \\
$f_\mathrm{scale}$ & Scaling parameter for fiber-loss correction & $\begin{cases}
\text{Clipped Normal}~[0.0, 2.0], \mu=1.0, \sigma=0.3 \\
\delta (1)
\end{cases}$ \\[8pt]
$f_\mathrm{run}^\mathrm{OB}$ & Fraction of young runaway stars        &
$\begin{cases}
\text{Uniform } [0.0, 1.0] \\
\text{Log Uniform } [-2.0, 0.0]
\end{cases}$ \\
\bottomrule\bottomrule
\end{NiceTabular}
\label{tab:priors}
\end{table*}

\begin{table}
    \centering
        \caption{Summary of the models tested in this work. Columns list the dust attenuation prescription (two or three components), the prior assumed on the runaway fraction $f_\mathrm{run}^\mathrm{OB}$, and any deviations from our base model.}
    \begin{NiceTabular}{l|c|c|c}[hvlines]
        \textbf{Model} & \textbf{Dust} & \textbf{Prior on $f_\mathrm{run}^\mathrm{OB}$} & \textbf{Notes} \\
        \hline
        \texttt{d2log20}      & 2   & Log-uniform $[-2, 0]$ & -- \\
        \texttt{d2log30}      & 2   & Log-uniform $[-3, 0]$ & -- \\
        \texttt{d2uni}        & 2   & Uniform $[0, 1]$      & -- \\
        \texttt{d2uniNoScale} & 2   & Uniform $[0, 1]$      & No $f_\mathrm{scale}$ \\
        \texttt{d3log20}      & 3 & Log-uniform $[-2, 0]$ & With $\mathrm{dust}_3$ \\
        \texttt{d3uni}        & 3 & Uniform $[0, 1]$      & With $\mathrm{dust}_3$ \\
        \texttt{d3log20NoD2}  & 3 (no $\tau_2$) & Log-uniform $[-2, 0]$ & With $\mathrm{dust}_3$ \\
    \end{NiceTabular}
    \label{tab:models}
\end{table}

\subsection{Uniform Prior Model}

The most straightforward assumption is to adopt a uniform prior for $f_\mathrm{run}^\mathrm{OB}$ (model \texttt{d2uni}). However, it is important to note that a uniform prior does not imply a lack of informative assumptions. In practice, the prior on the escape fraction is implicitly shaped by the choice of priors on the dust attenuation parameters, $\tau_2$. We repeat the setup of \texttt{d2uni} including the three component dust model ('\texttt{d3uni}').

\subsection{NoScale Model}

The NoScale model (\texttt{d2uniNoScale}) is identical to the \texttt{d2uni} setup, except that it omits the parameter $f_\mathrm{scale}$. By removing this parameter, \texttt{d2uniNoScale} isolates the influence of fiber-loss corrections on the inferred value of $f_\mathrm{esc}^\mathrm{LyC}$. We find that excluding $f_\mathrm{scale}$ leads to slightly poorer fits compared to \texttt{d2uni}, with the total $\chi^2$ increasing on average by a factor of 1.03, while the inferred $f_\mathrm{esc}^\mathrm{LyC}$ remains unaffected.

When focusing specifically on the $\chi^2$ contribution from the emission lines, the \texttt{d2uniNoScale} model yields an average value 1.3 times higher than that of the \texttt{d2uni} model. Based on this degradation in the emission line fit quality, we conclude that including $f_\mathrm{scale}$ as a free parameter is necessary and therefore retain it in our default model setup.

\subsection{Log-Uniform Prior Model}
\label{subsec:log20}

We additionally test a log-uniform prior for the parameter $f_\mathrm{run}^\mathrm{OB}$, spanning the range $10^{-2}$ to $1$ (model \texttt{d2log20}). This prior is motivated by the distribution of escape fractions inferred from the LzLCS sample and allows us to evaluate the sensitivity of our results to prior assumptions that favor lower escape fractions.

We then extend this setup by incorporating the three-component dust model, resulting in the \texttt{d3log20} configuration. Finally, we include a variant, \texttt{d3log20Nod2}, which is identical to \texttt{d3log20} but omits the second dust component (\texttt{dust2}). This configuration approximates a picket-fence scenario, where radiation from young stars is either fully attenuated by the birth-cloud dust or escapes without additional diffuse attenuation.

\begin{figure}
    \centering
    \includegraphics[width=\linewidth]{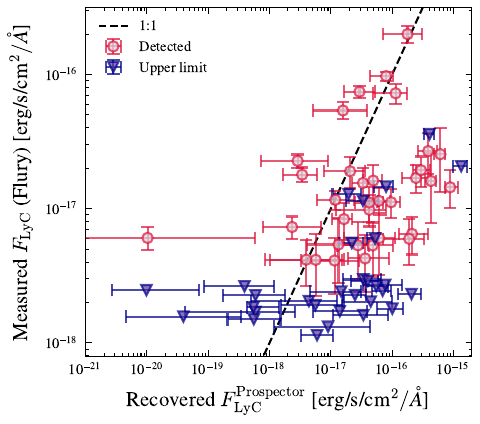}
    \caption{Comparison between the observed Lyman continuum fluxes from \citet{flury_low-redshift_2022} and the fluxes recovered with \texttt{Prospector} for the \texttt{d2log20} model. 
    We compare the measured $F_\mathrm{LyC}$ to the predicted $F_\mathrm{LyC}^\mathrm{Prospector}$, with orange circles indicating detections and blue triangles representing upper limits. Recovered fluxes were measured in the same variable wavelength ranges as in \citet{flury_low-redshift_2022}. Error bars indicate the 16th–84th percentile credible intervals, and the dashed line marks the one-to-one relation. \texttt{Prospector} systematically predicts lower fluxes than observed, suggesting a potential discrepancy between the measured line-of-sight flux and the inferred global flux.}
    \label{fig:lyman}
\end{figure}

\subsection{Performance Across Models}
\label{subsec:performanceAcrossModels}
Our models were unable to successfully fit the galaxies J090918+392925 and J081409+211459. We therefore remove them from all further discussion. All models successfully reproduce both photometry and emission lines, achieving a median photometric recovery of $\chi^2 \approx 3.5$ for the galaxy sample. An example fit to a galaxy’s photometry is provided in Figure~\ref{fig:fit_combined}, also in the Appendix.
We identify the best–fitting model for each galaxy using a hybrid criterion.  
We first compare the models via the Bayes factor; if the Bayes factor is non-decisive ($<5$), we then select the model that most accurately reproduces the photometry and emission lines, as quantified by the reduced $\chi^2$. All subsequent analysis is based on the resulting best-fit model for each galaxy.

To assess the reliability of our models, we perform a parameter recovery test, described in detail in Appendix~\ref{sec:parameterRecoveryTest} using our \texttt{d2uniNoScale} model. 
The test shows that the model systematically overpredicts $f_\mathrm{esc}^{\mathrm{LyC}}$, with a negative median pull $(f_\mathrm{esc,in}^\mathrm{LyC} - f_\mathrm{esc,out}^\mathrm{LyC}) \,/\,\sigma_{f_\mathrm{esc}}$ for $\log_{10}(f_\mathrm{esc}^\mathrm{LyC}) > -2.5$, indicating that the formal posterior uncertainties underestimate the true scatter in the downward direction. We therefore inflate the lower uncertainty by a factor of 1.5, after which the input values are recovered within $1\sigma$.
A key limitation arises in cases with very low escape fractions ($\log f_\mathrm{esc} < -2.5$) are overpredicted for dust-free systems. When $f_\mathrm{esc}^\mathrm{LyC}$ is low and dust is absent, nearly all ionizing photons are absorbed by the nebular gas, leading to very high ionization states. However, the nebular emission is modeled using a pre-computed grid that imposes an upper limit on the ionization parameter and is further constrained by a prior that disfavors extreme ionization states. As a result, the model cannot reproduce the observed emission line strengths under the correct physical conditions and instead compensates by overestimating the escape fraction.

Nevertheless, in cases where low escape fractions are not exclusively the result of strong H\,\textsc{i} absorption without any effect of dust attenuation, our framework can robustly distinguish between galaxies with $\log f_\mathrm{esc} \lesssim -2$ and galaxies with $\log f_\mathrm{esc} \gtrsim -0.5$. Finer distinctions are possible, but the corresponding posterior distributions begin to overlap.

Parameter estimates across our models are generally consistent, agreeing within $2\sigma$ for stellar mass, gas-phase metallicity, and, most importantly, $f_\mathrm{esc}^\mathrm{LyC}$. In Fig.~\ref{fig:compare_all} in the Appendix we show a comparison of the results by plotting the parameters against the results from our best performing model \texttt{d2uniNoScale}. The notable exception of the overall agreement is the \texttt{d3log20nod2} model, which systematically yields higher $f_\mathrm{esc}^\mathrm{LyC}$, with median values offset by nearly 2 dex compared to all other models. Unlike the other setups, \texttt{d3log20nod2} excludes a diffuse dust component while retaining separate dust screens for young and old stellar populations. Although the model reproduces the photometry and emission lines with comparable Bayesian evidence, its $\chi^2$ values are on average 50 units worse than the next best model.

\begin{figure}
    \centering
    \includegraphics[width=\linewidth]{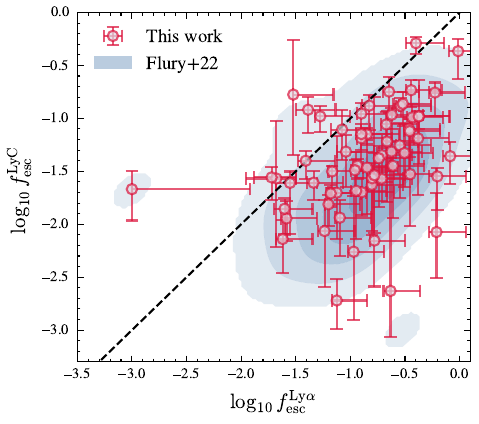}
    \caption{Comparison between $f_\mathrm{esc}^{\mathrm{Ly}\alpha}$ and the escape fraction inferred from SED fitting, $f_\mathrm{esc}^\mathrm{LyC}$ (orange circles), together with the $f_{\mathrm{esc, H}\beta}^\mathrm{LyC}$ proxy from \citet{flury_low-redshift_2022} shown as blue density contours. The black dashed line indicates the 1:1 relation. The highest $f_\mathrm{esc}^{\mathrm{Ly}\alpha}$ are well constrained by the 1:1 line, consistent with our SED-inferred $f_\mathrm{esc}^\mathrm{LyC}$ representing a global, angle-averaged escape fraction.}
    \label{fig:lyalpha}
\end{figure}

A comparison between the observed $F_{\mathrm{LyC}}$ and the model predictions is shown in Fig.~\ref{fig:lyman}. A log-rank test accounting for upper limits strongly rejects the hypothesis that the Prospector-predicted and observed LyC fluxes are drawn from the same parent distribution ($\chi^2 = 25.4, p = 4.6 \times 10^{-7}$). However, this does not necessarily indicate poor fits, as the observed LyC flux is inherently line-of-sight dependent, whereas the SED model yields a global, angle-averaged prediction. Following the line-of-sight interpretation discussed in \citet{flury_low-redshift_2022}, we find that the Prospector-predicted LyC flux generally acts as an upper bound to the measured LyC. This behavior can be understood in the context of a picket-fence ISM geometry: while the model prediction represents the total LyC escaping the galaxy, the observed flux probes a single line of sight, which may intersect partially or fully obstructed regions of the ISM. In a patchy medium characterized by small, optically thin channels \citep{saldana-lopez_low-redshift_2022,Flury:2025aa}, alignment between the line of sight and such channels is expected to be rare, such that in most cases $F_{\mathrm{LyC}} < F_{\mathrm{LyC}}^{\mathrm{pred}}$. Only when the line of sight intersects an optically thin channel would we expect $F_{\mathrm{LyC}} \gtrsim F_{\mathrm{LyC}}^{\mathrm{pred}}$. In Fig.~\ref{fig:lyalpha}, we compare the $f_\mathrm{esc}^{\mathrm{Ly}\alpha}$ estimates of \citet{flury_low-redshift_2022-1} with our inferred $f_\mathrm{esc}^\mathrm{LyC}$ and find that the latter forms an effective upper envelope for $f_\mathrm{esc}^{\mathrm{Ly}\alpha}$, again consistent with expectations from a picket-fence geometry.


\section{Results and Comparison to High-Redshift Galaxies}
\label{sec:results}

\begin{figure*}
    \centering
    \includegraphics[width=\linewidth]{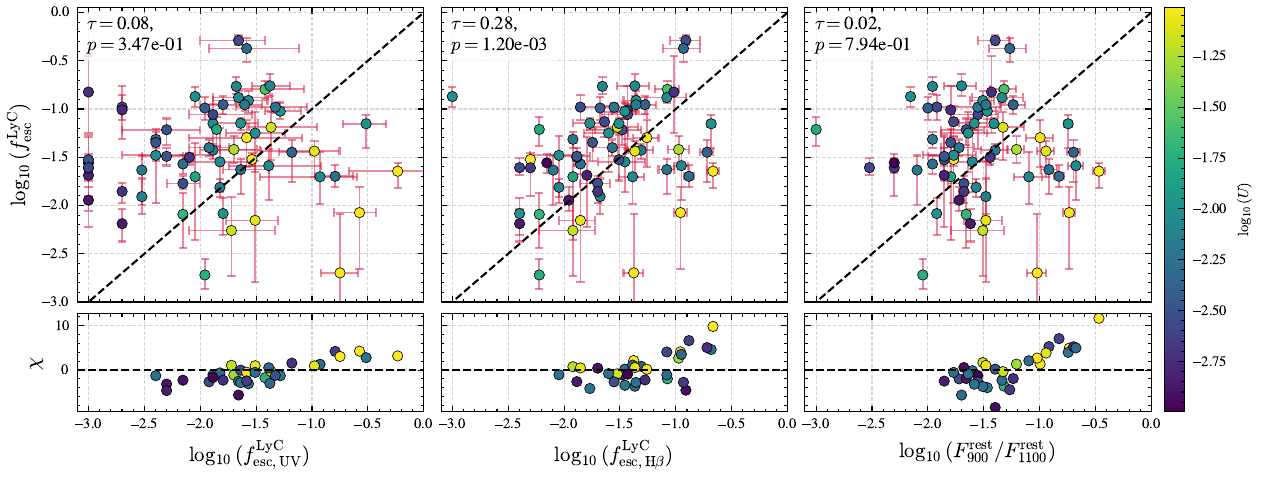}
    \caption{Comparison between the escape-fraction estimates $f_{\mathrm{esc,UV}}^{\mathrm{LyC}}$, $f_{\mathrm{esc,H}\beta}^{\mathrm{LyC}}$, and $F_{900}^{\mathrm{rest}}/F_{1100}^{\mathrm{rest}}$ from \citet{flury_low-redshift_2022} and the $f_{\mathrm{esc}}^{\mathrm{LyC}}$ inferred with \texttt{Prospector}. The $f_{\mathrm{esc}}^{\mathrm{LyC}}$ values presented here are derived using the restricted wavelength window described in Section~\ref{sec:data}. Each point represents a galaxy for which the best-fitting model was selected using a hybrid evidence--$\chi^2$ criterion.
    The dashed black line marks the one-to-one relation. In the corner of each panel, we report Kendall’s $\tau$ rank correlation coefficient and the corresponding $p$-value between the escape fraction estimations. Among the three diagnostics considered, $f_{\mathrm{esc,H}\beta}^{\mathrm{LyC}}$ is the only quantity from \citet{flury_low-redshift_2022} that exhibits a statistically moderate but significant correlation with our inferred escape fractions, with $\tau = 0.28$ and $p = 10^{-3}$. The points are colored by the ionization parameter $\log(U)$ inferred from the SED fit. While the escape fraction estimates reported by \citet{flury_low-redshift_2022} show moderate to strong correlations with $\log(U)$, with Kendall’s $\tau = 0.50$, $0.09$, and $0.23$ for $f_{\mathrm{esc,UV}}^{\mathrm{LyC}}$, $f_{\mathrm{esc,H}\beta}^{\mathrm{LyC}}$, and $F_{900}^{\mathrm{rest}}/F_{1100}^{\mathrm{rest}}$, respectively, we find virtually no correlation between the restricted $f_{\mathrm{esc}}^{\mathrm{LyC}}$ inferred in this work and $\log(U)$, with $\tau = -0.07$}.
    \label{fig:fesc}
\end{figure*}

\begin{table*}
\centering
    \caption{Subset of escape fractions derived from SED modelling for each galaxy. For each object we list the median and the 16th--84th percentile credible intervals for the absolute escape fraction of ionizing photons $f_{\mathrm{esc}}^{\mathrm{LyC}}$ (derived by integrating the full bestfit model spectra at $\lambda_\mathrm{rest}$<912\AA) found by the respective best model according to the hybrid evidence~+\,$\chi^2$ selection described in Section~\ref{subsec:performanceAcrossModels}, and the restricted version of the escape fraction obtained from a $20~$\AA$~$region around the central rest-frame wavelength ($\lambda_\mathrm{central}^\mathrm{rest}$). The lower uncertainties have been inflated by a factor of 1.5 to account for systematic scatter identified in the parameter recovery test (Section~\ref{sec:parameterRecoveryTest}).
    }
\label{tab:escape_fractions}
\begin{tabularx}{\textwidth}{XYYY}
\toprule
ID & $f_{\mathrm{esc}}^{\mathrm{LyC}}$ & $f_{\mathrm{esc, restricted}}^{\mathrm{LyC}}$ & $\lambda_\mathrm{central}^\mathrm{rest}$ \\[3pt]
\midrule
J003601+003307 & $(3.57^{+0.50}_{-0.99})\times 10^{-2}$ & $(3.00^{+0.48}_{-0.44})\times 10^{-2}$ & 860 \\[3pt]
J004743+015440 & $(4.49^{+0.72}_{-1.69})\times 10^{-2}$ & $(3.75^{+0.64}_{-0.69})\times 10^{-2}$ & 860 \\[3pt]
J011309+000223 & $(5.13^{+0.67}_{-1.20})\times 10^{-1}$ & $(5.13^{+0.67}_{-0.53})\times 10^{-1}$ & 890 \\[3pt]
J012217+052044 & $(1.75^{+0.38}_{-0.89})\times 10^{-1}$ & $(1.60^{+0.37}_{-0.33})\times 10^{-1}$ & 850 \\[3pt]
J012910+145935 & $(2.69^{+0.67}_{-1.31})\times 10^{-2}$ & $(2.69^{+0.67}_{-0.58})\times 10^{-2}$ & 890 \\[3pt]
\bottomrule
\end{tabularx}
\end{table*}

\begin{figure}
    \centering
    \includegraphics[width=\linewidth]{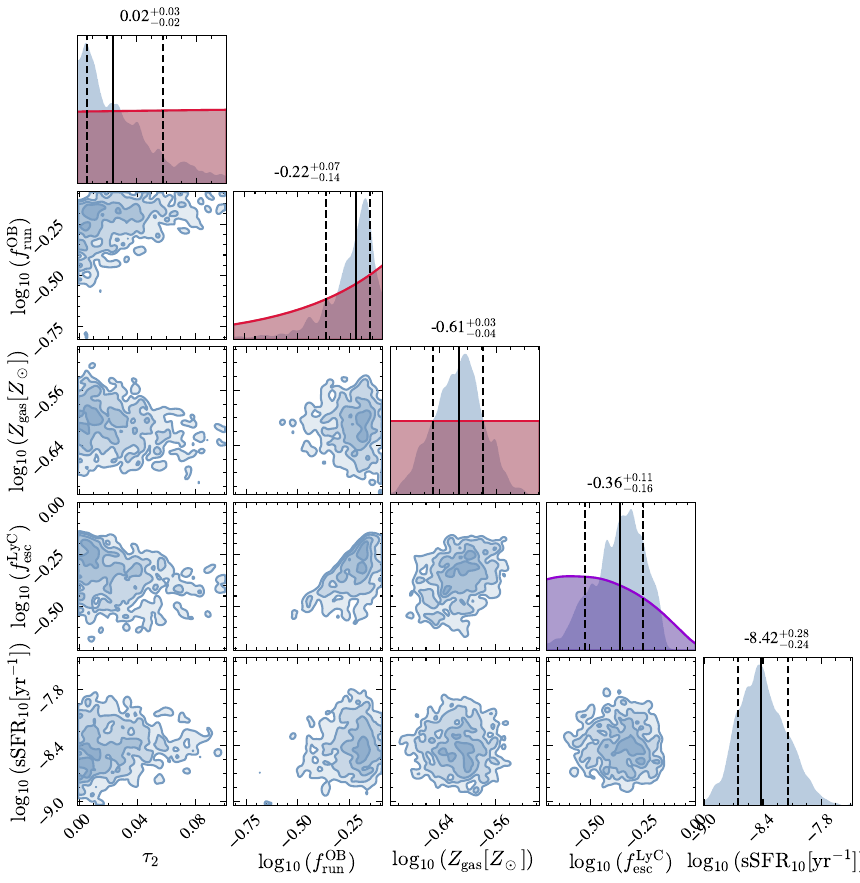}
    \caption{Posterior distributions (blue) for the best-fit model of galaxy J164849+495751, which exhibits a high escape fraction, $f_\mathrm{esc}^\mathrm{LyC} \sim 35_{-16}^{+13}\%$. Diagonal panels show the marginalized one-dimensional distributions, with vertical lines marking the median and 16th–84th percentiles, while off-diagonal panels display two-dimensional parameter correlations. Shown are the directly fitted parameters (diffuse dust optical depth, $f_\mathrm{run}^\mathrm{OB}$, and gas-phase metallicity) together with their priors (orange), as well as the derived quantities $\log_{10}(f_\mathrm{esc}^\mathrm{LyC})$, with its induced prior (purple), and $\log_{10}\mathrm{sSFR}{10}$. The comparison between posteriors and priors indicates that the inferred parameters are constrained by the data.}
    \label{fig:corner_small}
\end{figure}

In this section, we present the $f_\mathrm{esc}^\mathrm{LyC}$  inferred from our models and place them in context by comparing the physical properties of the LzLCS galaxies to the properties of high-redshift EoR (z>5) populations. We focus on three key aspects that are particularly relevant for understanding galaxies in the EoR: SFHs (Section~\ref{sec:sfr}), gas-phase enrichment (Section~\ref{sec:gasPhaseMetal}), and the size–mass relation (Section~\ref{sec:size}). 
Having established our modelling framework and validated its robustness in the previous subsection, we now present the key results obtained for individual galaxies. The $f_\mathrm{esc}^\mathrm{LyC}$ results are summarized in Tab.~\ref{tab:escape_fractions}. For each galaxy, we report the value from its individual best-fitting model and from SEDs restricted to the parameter range used by \citet{flury_low-redshift_2022-1}, based on their LyC measurements avoiding contamination. Using \texttt{d2uniNoScale} as a reference, we find that the $f_\mathrm{esc}^\mathrm{LyC}$ estimates from all other models except \texttt{d3lognod2} are highly consistent with each other, showing strong rank and linear correlations (Kendall's $\tau \simeq 0.74\text{–}0.88$) and near-complete $1\sigma$ interval overlap. In contrast, \texttt{d3lognod2} exhibits poor correlation with our other models, large systematic offsets, and significant disagreement for the majority of galaxies ($\sim 60\%$ exceeding $2\sigma$). The \texttt{d3lognod2} configuration assumes an extreme, dust-free ionizing photon budget in which all LyC photons not producing nebular emission escape the galaxy. While this scenario may apply to rare systems, it is inconsistent with the majority of galaxies in our sample and leads to systematically higher $f_\mathrm{esc}^\mathrm{LyC}$. Accordingly, this model is strongly disfavored and is preferred for only one out of 66 galaxies. Comparing $f_\mathrm{esc}^\mathrm{LyC}$ to the restricted $f_\mathrm{esc, res}^\mathrm{LyC}$, we find that the inferred $f_\mathrm{esc, res}^\mathrm{LyC}$ are on average 0.08~dex lower when using the restricted range, thus smaller than the uncertainties connected to our estimations of significant $f_\mathrm{esc}^\mathrm{LyC}$.

In Figure~\ref{fig:fesc}, we compare our inferred $f_\mathrm{esc}^\mathrm{LyC}$ values to the different escape fraction estimates of \citet{flury_low-redshift_2022-1}, $f_\mathrm{esc,~UV}^\mathrm{LyC},~f_{\mathrm{esc,~H}\beta}^\mathrm{LyC}$ and $F_{900}^{\mathrm{rest}}/F_{1100}^{\mathrm{rest}}$. Overall, our analysis yields systematically lower $f_\mathrm{esc}^\mathrm{LyC}$ than the estimations by \citet{flury_low-redshift_2022-1}. We find the strongest correlation with the H$\beta$-based escape fractions (Kendall's $\tau \simeq 0.28 \pm 0.1$, corresponding to a significance of $3.0 \sigma$). The correlation with UV-based escape fractions and the empirical UV flux ratio $\log(F_{900}^\mathrm{rest}/F_{1100}^\mathrm{rest})$ are insignificant, (Kendall's $\tau \simeq 0.08 \pm 0.1$, $0.4 \sigma$ and Kendall's $\tau \simeq 0.02 \pm 0.09$), consistent with the expectation that this ratio probes a strongly line-of-sight–dependent quantity and therefore does not directly trace the global LyC escape fraction inferred from SED fitting. We further find that, in the analysis of \citet{flury_low-redshift_2022-1}, galaxies with higher inferred LyC escape fractions also tend to exhibit higher ionization parameters $\log U$ as determined in our SED fitting. We find weak to strong positive correlation between the Flury escape-fraction estimates and $\log U$, with Kendall’s rank correlation coefficients of $\tau \simeq 0.5 \pm 0.09$ ($5.68\,\sigma$) for $f_{\mathrm{esc,UV}}^{\mathrm{LyC}}$, $\tau \simeq 0.09 \pm 0.09$ ($0.6\,\sigma$) for $f_{\mathrm{esc,H}\beta}^{\mathrm{LyC}}$, and $\tau \simeq 0.23 \pm 0.09$ ($2.5\,\sigma$) for $F_{900}^{\mathrm{rest}}/F_{1100}^{\mathrm{rest}}$.

This likely arises because their methods do not fit for $\log U$, allowing part of its influence to be absorbed into the inferred $f_\mathrm{esc}^\mathrm{LyC}$. In our framework, $\log U$ is treated as a free parameter, breaking this degeneracy; as a result, our recovered $f_\mathrm{esc}^\mathrm{LyC}$ values show no significant dependence ($\tau \simeq -0.07\pm0.09$) on the ionization state in contrast to the literature \citep{Jaskot:2013aa, Izotov2018MNRAS.478.4851I, flury_low-redshift_2022}.
Figure~\ref{fig:corner_small} presents a subset of the posterior distributions for key parameters of an example galaxy J164849+495751 for which we find $f_\mathrm{esc}^\mathrm{LyC} = 43_{-12}^{+12} \%$. The full set of posteriors is provided in Appendix~\ref{fig:corner_highfesc}. When examining the posterior distributions for this galaxy, we recover the expected degeneracies between the SFH parameters and the stellar mass, as well as between stellar metallicity and stellar age. We also identify mild degeneracies between $f_\mathrm{esc}^\mathrm{LyC}$ and $f_\mathrm{scale}$. Physically, a poor constraint on $f_\mathrm{scale}$ could lead to a misattribution of escaping ionizing photons to fiber losses or vice versa. However, the degeneracy is sufficiently weak that it does not significantly impact our inference of $f_\mathrm{esc}^\mathrm{LyC}$ beyond what is already captured by the posterior uncertainty. Additionally, $f_\mathrm{scale}$ is not included in the best performing model which accounts for 59 of the 64 galaxies.

Next, we place our results in a broader context by comparing the LzLCS galaxies to both local and high-redshift populations, focusing on three fundamental properties: star formation activity (Section~\ref{sec:sfr}), gas-phase metallicities (Section~\ref{sec:gasPhaseMetal}), and sizes (Section~\ref{sec:size}). 
\subsection{Star Formation Histories}
\label{sec:sfr}

\begin{figure}
    \centering
    \includegraphics[width=\linewidth, alt={The median star formation history rises monotonically from the second-oldest time bin to the present. The shaded percentile region remains relatively constant in width, suggesting that most galaxies exhibit a similar rising trend toward recent star formation.}]{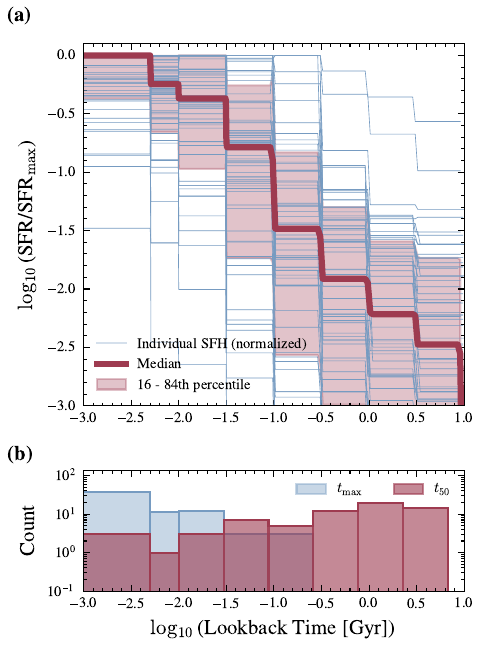}
    \caption{\textbf{(a)} Stacked star formation histories of the LzLCS galaxies. Each individual SFH, normalized to its maximum, is shown as a thin blue line. The thick orange line marks the population median, with the shaded region denoting the 16th–84th percentile range. The steadily rising median toward recent times, together with the narrow percentile spread, demonstrates that nearly all galaxies in the sample are experiencing a strong, recent ($<10$ Myr) burst of star formation.  
    \textbf{(b)} Distributions of the half-mass lookback time ($t_{50}$; when 50\% of the stellar mass was formed) and the time of peak star formation ($t_{\mathrm{max}}$). Most galaxies have $t_{50}$ within the last $\sim$2 Gyr, while their peak activity lies within the past 10 Myr. This confirms that the LzLCS sample represents an intensely star-forming population dominated by very recent bursts.}
    \label{fig:sfh_median}
\end{figure}

Figure~\ref{fig:sfh_median} shows the inferred star formation histories (SFHs) and the corresponding mass-weighted and burst ages for the LzLCS galaxies. All systems are currently in a phase of elevated star formation activity, with SFHs that rise steeply toward recent times. For most galaxies, the half-mass assembly time $t_{50}$ occurs within the last $\sim 2$~Gyr, while the peak of star formation typically lies within the past $\sim 10$~Myr. Older $t_{50}$ values cannot be robustly excluded due to the strong outshining of older stellar populations by the ongoing starburst.

Quantitatively, we find that the inferred LyC escape fractions exhibit moderate to weak but statistically significant anti-correlations with the absolute star formation rate averaged over different timescales, including $\mathrm{sSFR}_{100}$ and $\mathrm{sSFR}_{10}$ (Kendall's $\tau \simeq -0.28$ and $\tau \simeq -0.21$) and weaker with shorter timescale averages such as $\mathrm{SFR}_5$ and $\mathrm{SFR}_{10}$ (Kendall's $\tau \simeq -0.177$ and $\tau \simeq -0.176$). In contrast, surface-density–based quantities show no statistically significant correlation with $f_\mathrm{esc}^\mathrm{LyC},~~\tau < 0.168$.
These results are difficult to interpret because the posterior distributions of the fits exhibit degeneracies between $f_\mathrm{esc}^\mathrm{LyC}$ and recent changes in the star formation rate. The number of ionizing photons produced —approximately proportional to the instantaneous SFR— can either ionize neutral hydrogen and power recombination lines such as H$\alpha$, or escape the galaxy. As a result, an anticorrelation between $f_\mathrm{esc}^\mathrm{LyC}$ and recent star formation indicators is expected.
Table~\ref{tab:sfh} shows the inferred parameters relating to the SFH of our galaxies.

\begin{figure}
    \centering
    \includegraphics[width=\linewidth]{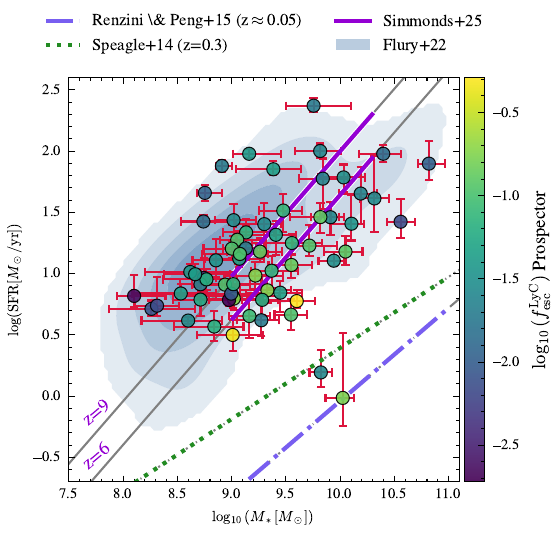}
    \caption{Stellar mass versus star formation rate for the LzLCS sample.  SFRs averaged over the past 10 Myr ($\mathrm{SFR}_{10}$) from our \texttt{Prospector} fits are shown as circles, colored by their $f_\mathrm{esc}^\mathrm{LyC}$. For comparison, estimates from \citet{flury_low-redshift_2022} based on FUV are shown as blue contours. Literature star-forming main sequence (SFMS) relations are overplotted: local relations from \citet{renzini_objective_2015} (green dotted; $z\approx 0.05$) and \citet{speagle_highly_2014} (blue dash-dotted; $z\approx0.3$), as well as high-redshift relations based on \textit{JWST} observations at $z=6$ and $z=9$ from \citet{simmonds_impact_2024}. 
    Extrapolations of published relations beyond their calibrated mass ranges are indicated by thin gray extensions of the colored lines. The LzLCS galaxies align more closely with the high-redshift SFMS than with local relations.}
    \label{fig:sfms}
\end{figure}

In Figure~\ref{fig:sfms}, we compare the recent SFRs of the LzLCS galaxies to local star-forming systems, high-redshift populations observed with \textit{JWST}, and to the values previously reported for this sample by \citet{flury_low-redshift_2022-1}. Our stellar mass estimates are on average $\sim0.35$~dex higher than those reported by \citet{flury_low-redshift_2022-1}, who fitted for the stellar mass using \texttt{Prospector} and the photometry from SDSS and GALEX. This offset likely reflects differences in the adopted modeling assumptions: by fitting simultaneously for additional parameters such as stellar metallicity, our framework partially breaks the classical age–metallicity degeneracy \citep{Worthey1994Comprehensive, 2012MNRAS.422.3285P}, which can otherwise bias mass estimates. As a consistency check, we nevertheless find a strong and highly significant rank correlation between the two mass estimates (Kendall's $\tau \simeq 0.57$, $p \sim 6\times10^{-11}$), indicating that the relative ordering of stellar masses is well preserved despite the systematic offset. Despite these mass differences, our $\mathrm{SFR}{10}$ values are in excellent agreement with independent star-formation tracers. In particular, we find a strong and highly significant correlation between $\mathrm{SFR}{10}$ and SFRs inferred from UV fluxes using the calibration of \citet{jr_star_2012} reported by \citet{flury_low-redshift_2022-1} ($\tau \simeq 0.38$, $p \sim 1\times10^{-5}$), validating the recovery of recent star formation in our SED fits.

Relative to the local star-forming main sequence (SFMS) at $z\approx0.05$ \citep{renzini_objective_2015} and $z\approx0.3$ \citep{speagle_highly_2014}, the LzLCS galaxies exhibit systematically elevated SFRs. We find an average offset of $1.2$~dex for galaxies with $\log(M_\ast/M_\odot)<9$ and $0.7$~dex for galaxies above $\log(M_\ast/M_\odot)>10$, confirming that the LzLCS sample lies well above the local SFMS. The LzLCS galaxies align much more closely with high-redshift relations: for example, the recent JWST-based SFMS from \citet{simmonds_bursting_2025} at $z=6$ and $z=9$. In this case, the average offset is only $0.05$~dex for galaxies with $9 < \log(M_\ast/M_\odot) < 10$, though the most massive systems remain around $\sim0.5$~dex below the high-redshift relation.


\subsection{Gas-Phase Enrichment}
\label{sec:gasPhaseMetal}

We find that the gas-phase metallicities in our sample are systematically higher than the stellar metallicities, with an average offset of 0.67~dex. The stellar metallicities are subject to considerable uncertainty, primarily due to the limited constraints from absorption features in our data. Furthermore, the two metallicity indicators trace different chemical elements: gas-phase metallicities primarily reflect the abundance of $\alpha$-elements such as oxygen, while stellar metallicities are sensitive to a combination of $\alpha$- and iron-peak elements. The relative contribution of these elements depends on the wavelength regime; stellar metallicities derived from optical spectra are sensitive to both $\alpha$- and Fe-group elements, whereas UV absorption features tend to be more strongly dominated by iron. As a result, gas-phase metallicities are expected to exceed stellar metallicities by a factor of $\sim$2--5 \citep{strom_chemical_2022, arellano-cordova_first_2022}.

\begin{figure}
    \centering
    \includegraphics[width=\linewidth]{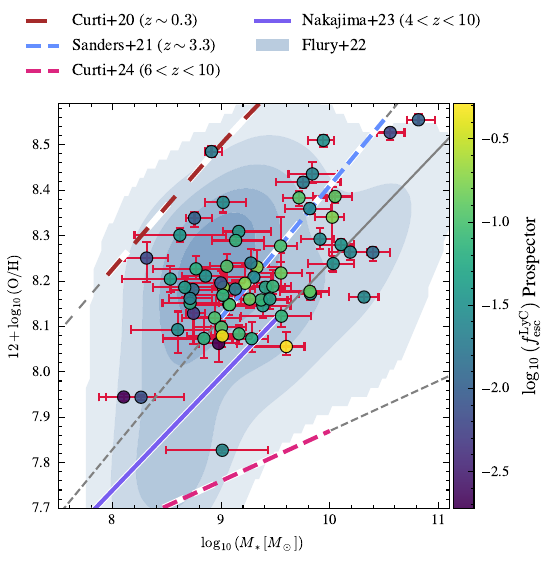}
    \caption{Comparison of the mass–metallicity relation across redshifts with the newly derived gas-phase metallicities and masses for the LzLCS sample, shown as circles, colored by their $f_\mathrm{esc}^\mathrm{LyC}$. Values derived for the LzLCS by \citealt{flury_low-redshift_2022} are shown as blue contours. For local galaxies at $z\approx 0$, we include SDSS (brown dashed line; \citealt{curti_massmetallicity_2020}) and electron-temperature-based metallicities (dark orange solid line; \citealt{yates_present-day_2020}). At intermediate redshift, we show MOSDEF at $z\sim 3.3$ (light blue dotted line; \citealt{sanders_mosdef_2021}), which is in closest agreement with our LzLCS results. At higher redshifts, we compare to gas-phase metallicities from \textit{JWST} observations: JADES $3<z<10$ (pink dashed line; \citealt{curti_jades_2024}) and galaxies at $4<z<10$ (blue solid line; \citealt{nakajima_jwst_2023}), both of which predict significantly lower metallicities. 
    Extrapolations of published relations beyond their calibrated mass ranges are shown as thin gray extensions of the colored lines. Although the LzLCS galaxies fall below the local gas-phase metallicity relation, they show systematically higher metallicities than galaxies at high redshift, suggesting they may not be fully representative of typical reionization-era systems. Systems with the higher $f_\mathrm{esc}^\mathrm{LyC}$ of the LzLCS exhibit low metallicity compared to their mass.}
    \label{fig:zgas_mass}
\end{figure}

In Figure~\ref{fig:zgas_mass}, we place our estimations in the context of the mass-metallicity relation (MZR) across cosmic time and compare them to estimates from \citet{flury_low-redshift_2022-1}. We compare our gas-phase metallicities with the electron-temperature–based values derived from the [O,\textsc{iii}]~$\lambda\lambda4363,4959,5007$ emission lines and find weak but non-negligible agreement. The Monte Carlo Kendall analysis yields $\tau \simeq 0.16 \pm 0.06$ ($p \simeq 0.07$), while the classical Kendall statistic gives $\tau \simeq 0.19$ ($p \simeq 0.03$). This level of correlation indicates broad consistency in relative gas-phase enrichment, albeit with substantial scatter, as expected given the different physical assumptions and sensitivities of SED-based and direct $T_e$-based metallicity estimates.
Relative to the local $z \sim 0.1$ relation from SDSS \citep{curti_massmetallicity_2020}, the LzLCS galaxies are clearly offset toward lower metallicities: by $-0.28$~dex at $\log (M_\ast/M_\odot) < 9$, $-0.38$~dex at $9 < \log (M_\ast/M_\odot) < 10$, and $-0.40$~dex at $\log (M_\ast/M_\odot) > 10$. In contrast, the agreement with intermediate-redshift galaxies ($z \sim 3.3$; \citealt{sanders_mosdef_2021}) is significantly better, with much smaller offsets of $+0.10$, $-0.02$, and $-0.13$~dex in the same mass bins. 
The comparison with JWST observations at $4 < z < 10$ from \citet{nakajima_jwst_2023} shows an intermediate level of agreement: although still offset from the $z > 6$ population in the lower mass bins (by $+0.22$ and $+0.13$~dex), the highest-mass galaxies in our sample ($\log (M_\ast/M_\odot) > 10$) are nearly consistent, with an average offset of only $+0.05$~dex, but the metallicity distribution of these galaxies exhibits a significantly larger scatter than in the lower mass bins.
 
These results suggest that while the LzLCS galaxies are clearly metal-poor compared to local star-forming galaxies, they are more consistent with the chemical properties of intermediate-redshift populations, and do not reach the very low metallicities seen in galaxies at $z > 6$.

\subsection{Size–Mass Relation}
\label{sec:size}

\begin{figure}
    \centering
    \includegraphics[width=\linewidth]{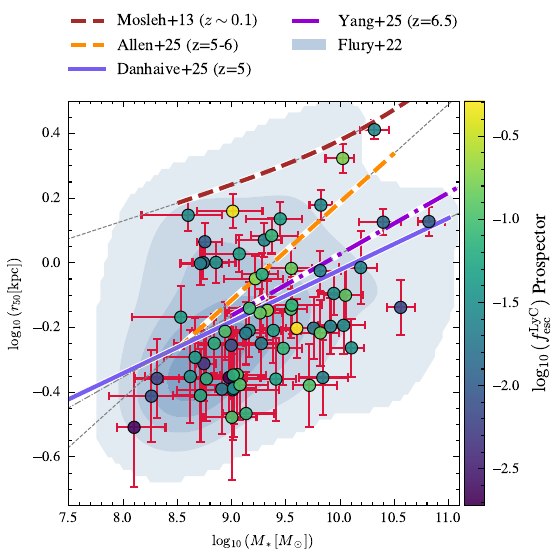}
    \caption{Size–mass relation of the LzLCS galaxies, where sizes ($r_{50}$) are measured in the NUV \citep{flury_low-redshift_2022} and stellar masses are re-derived in this work. Our estimations for the LzLCS sample are shown as circles, colored by their $f_\mathrm{esc}^\mathrm{LyC}$, the ones derived by \citet{flury_low-redshift_2022} as blue contours. For comparison, we show the local relation from SDSS galaxies at $z\sim 0.1$ (brown dashed line; \citealt{mosleh_robustness_2013}), high-redshift estimations from \textit{JWST}: at $z=5$–6 (green dotted line; \citealt{allen_galaxy_2025}),  at $z=5$ (blue solid line; \citealt{2025arXiv251006315D}), and COSMOS-Web at $z=6.5$ (dark violet dash-dotted line; \citealt{yang_cosmos-web_2025}). 
    Extrapolations of published relations beyond their calibrated mass ranges are indicated by thin gray extensions of the colored lines. The LzLCS galaxies are systematically more compact than the local SDSS population, aligning more closely with the size–mass relation observed at high redshift. While some systems with elevated $f_\mathrm{esc}^\mathrm{LyC}$ appear more compact than both local and high-redshift counterparts, our most reliable LyC leaker does not exhibit extreme compactness and instead lies closest to the local size–mass relation.}
    \label{fig:size_mass}
\end{figure}

The escape fraction $f_\mathrm{esc}^\mathrm{LyC}$ has been suggested to correlate with galaxy compactness \citep{flury_low-redshift_2022-1, kim_small_2023}. To investigate this connection, we compare size estimates based on NUV flux from \citet{flury_low-redshift_2022-1} with our stellar mass estimates. In Figure~\ref{fig:size_mass}, we place the LzLCS galaxies on the size–mass relation and compare them with local star-forming galaxies from SDSS at $z\sim0.1$ \citep{mosleh_robustness_2013}. Across all mass bins, the LzLCS galaxies are significantly more compact, with mean offsets of $-0.52$~dex for $\log (M_\ast[M_\odot]) < 9$, $-0.52$~dex for $9 < \log (M_\ast[M_\odot]) < 10$, and $-0.50$~dex for $\log (M_\ast[M_\odot]) > 10$ relative to the SDSS relation. 

We further compare to recent JWST size measurements at $z\sim5$ \citep{Danhaive:2025aa}, finding that the LzLCS galaxies remain systematically more compact, with average offsets of $-0.09$, $-0.13$, and $-0.12$~dex across the same mass ranges. We do not find a clear correlation between $f_\mathrm{esc}^\mathrm{LyC}$ and galaxy size within the LzLCS sample, likely because it is already biased toward highly compact systems.

\begin{table*}
\centering
\caption{Stellar population parameters inferred for our galaxy sample. Columns list the median and 16th--84th percentile credible intervals of the stellar mass $\log_{10}(M_\ast/\mathrm{M_\odot})$, star formation rates averaged over the past 5, 10, and 100~Myr ($\mathrm{SFR}_{5}$, $\mathrm{SFR}_{10}$, $\mathrm{SFR}_{100}$), the mass-weighted stellar age $t_{50}$, the stellar metallicity $\log_{10}(Z_\ast/Z_\odot)$, and the gas-phase metallicity $\log_{10}(Z_\mathrm{gas}/Z_\odot)$. All values are derived from SED fitting.}
\label{tab:sfh}
\begin{tabularx}{\textwidth}{XYYYYYYR}
\toprule
ID & $\log_{10}(M_{\ast})$ & $\mathrm{SFR}_{5}$ & $\mathrm{SFR}_{10}$ & $\mathrm{SFR}_{100}$ & $t_{50}$ & $\log_{10}(Z_{\ast})$ & $\log_{10}(Z_\mathrm{gas})$ \\[3pt]
 & [$M_\odot$] & [$M_\odot~\mathrm{yr}^{-1}$] & [$M_\odot~\mathrm{yr}^{-1}$] & [$M_\odot~\mathrm{yr}^{-1}$] & [Gyr] & [$Z_\odot$] & [$Z_\odot$] \\[3pt]
\midrule
J003601+003307 & $8.62_{-0.42}^{+0.42}$ & $20.04_{-2.78}^{+3.04}$ & $10.30_{-1.40}^{+1.56}$ & $1.21_{-0.17}^{+0.20}$ & $1.48_{-1.48}^{+2.70}$ & $-1.13_{-0.23}^{+0.17}$ & $-0.39_{-0.02}^{+0.02}$ \\[3pt]
J004743+015440 & $9.38_{-0.34}^{+0.25}$ & $127.77_{-14.04}^{+23.27}$ & $71.02_{-7.97}^{+10.43}$ & $8.95_{-1.33}^{+2.77}$ & $0.53_{-0.52}^{+2.17}$ & $-0.95_{-0.16}^{+0.17}$ & $-0.54_{-0.03}^{+0.02}$ \\[3pt]
J011309+000223 & $9.60_{-0.32}^{+0.17}$ & $8.68_{-1.74}^{+2.79}$ & $5.96_{-0.74}^{+1.28}$ & $2.55_{-0.91}^{+0.70}$ & $4.46_{-2.22}^{+0.85}$ & $-0.42_{-0.52}^{+0.26}$ & $-0.63_{-0.02}^{+0.03}$ \\[3pt]
J012217+052044 & $9.33_{-0.25}^{+0.22}$ & $8.33_{-2.62}^{+3.22}$ & $7.30_{-1.74}^{+1.93}$ & $2.60_{-0.76}^{+0.99}$ & $2.31_{-1.87}^{+2.09}$ & $-0.98_{-0.20}^{+0.21}$ & $-0.46_{-0.04}^{+0.04}$ \\[3pt]
J012910+145935 & $9.91_{-0.18}^{+0.13}$ & $19.32_{-6.29}^{+7.41}$ & $28.84_{-9.59}^{+8.01}$ & $11.01_{-2.58}^{+3.75}$ & $2.98_{-2.02}^{+1.63}$ & $-1.35_{-0.18}^{+0.12}$ & $-0.40_{-0.02}^{+0.03}$ \\[3pt]
\bottomrule
\end{tabularx}
\end{table*}


\section{Symbolic and Linear Regression}
\label{sec:symreg}

\begin{figure*}
    \centering
    \includegraphics[width=\linewidth]{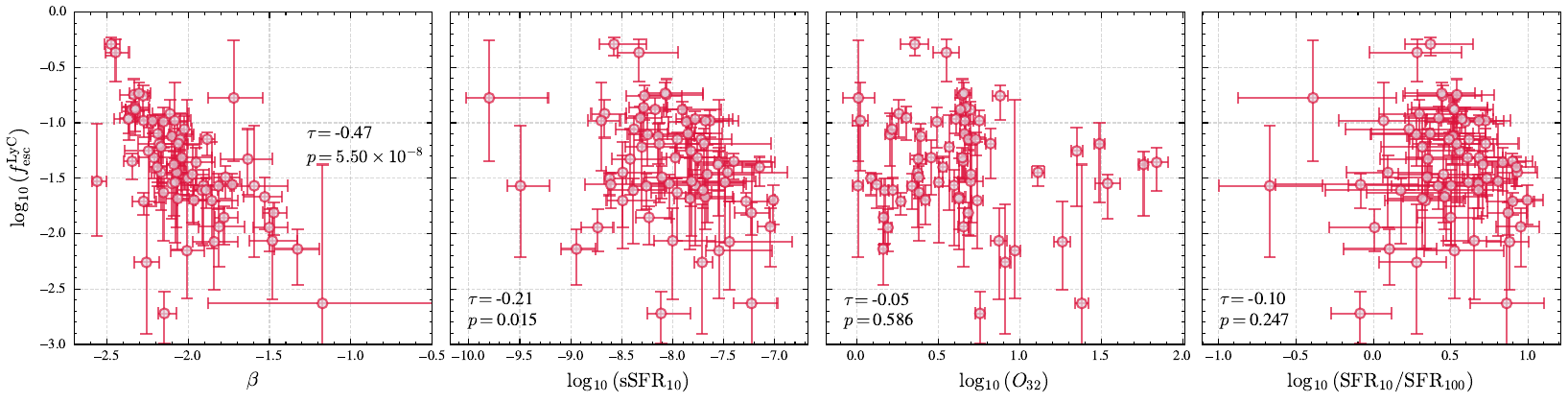}
    \caption{Comparison between the Lyman continuum escape fraction and other parameters: the UV continuum slope $\beta$, the specific star formation rate s($\mathrm{SFR}_{10}$), $O_{32}$, and the burstiness, defined as $\mathrm{SFR}_{10}/\mathrm{SFR}_{100}$.Each panel lists the classical Kendall's $\tau$ correlation coefficient quantifying the relationship between the plotted parameter and $\log(f_\mathrm{esc}^\mathrm{LyC})$.We find no significant correlations, except for the known $\beta$–$f_\mathrm{esc}^\mathrm{LyC}$ relation. This motivates the use of symbolic regression to search for more complex correlations.}
    \label{fig:placeholder}
\end{figure*}

We examine potential direct correlations between $f_\mathrm{esc}^\mathrm{LyC}$ and a set of global galaxy properties, a summary can be found in Tab.~\ref{tab:tau_results}. In Figure~\ref{fig:placeholder}, we show the escape fraction as a function of a subset of the properties we compare it to; the UV continuum slope $\beta$ (measured between $1250\lambda_\mathrm{rest} - 2500\lambda_\mathrm{rest}$), the star formation rate surface density, the specific star formation rate surface density, and their burstiness ($\mathrm{SFR}_{10}/\mathrm{SFR}_{100}$). Visual inspection and simple linear fits reveal no significant correlations with $f_\mathrm{esc}^\mathrm{LyC}$, apart from the well-established $\beta$–$f_\mathrm{esc}$ relation. This lack of straightforward trends motivates the use of symbolic regression (SR) to perform a data-driven search for more complex, potentially non-linear combinations of high-redshift observables that may serve as predictors of $f_\mathrm{esc}^\mathrm{LyC}$. 
Symbolic regression is a supervised learning technique that searches the space of analytical expressions to identify mathematical formulas that best describe a dataset. Unlike flexible machine-learning regressors that prioritize predictive performance over interpretability, symbolic regression produces closed-form expressions that can be directly interpreted and compared to theoretical expectations. We apply SR with the Python–Julia package \texttt{PySR} \citep{cranmer_discovering_2020}. \texttt{PySR} combines evolutionary search strategies with sparse regression and symbolic manipulation. Internally, it uses a graph neural network (GNN), which is guided towards the search of low-dimensional representations of the data, from which candidate equations are extracted and optimized.

\begin{table*}
\centering
\caption{Kendall $\tau$ rank correlations between the inferred LyC escape fraction ($f_{\mathrm{esc}}^{\mathrm{LyC}}$) and a range of physical diagnostics. 
The central correlation coefficient $\tau$ is computed using the median parameter estimates. 
Uncertainties $\sigma_{\tau,\mathrm{unc}}$ are obtained by propagating uncertainties through Monte-Carlo realizations, while $\sigma_{\tau,\mathrm{samp}}$ quantifies sample variance estimated via bootstrap resampling of the galaxy sample. 
The total uncertainty $\sigma_{\tau,\mathrm{tot}}$ is computed by adding these contributions in quadrature. 
p-values and corresponding Gaussian-equivalent significances $\sigma$ are reported for the correlations evaluated on the central values.For a sample size of n = 64, the critical value for a two-sided test at the $\alpha$ = 0.05 level is $|\tau| \ge 0.168$, below which the null hypothesis of no monotonic association cannot be rejected. Rows are shaded from dark to light gray to indicate decreasing levels of statistical evidence: extremely statistically significant, highly significant, significant at the 1\% level, and significant at the 5\% level. Correlations in all remaining rows are not statistically significant at the 5\% threshold.}
\label{tab:tau_results}
\begin{tabularx}{\textwidth}{X|RRY|RY|Y}
\toprule
Parameter & $\tau_0$ & p-value & $\sigma$ & $\sigma_{\tau,\mathrm{meas}}$ & $\sigma_{\tau,\mathrm{samp}}$ & $\sigma_{\tau,\mathrm{tot}}$ \\[3pt]
\midrule
\rowcolor{gray!40}
$\beta_{1200}$ & -0.534 & $4.55\times10^{-10}$ & 6.124 & 0.050 & 0.079 & 0.094 \\[3pt]
\rowcolor{gray!40}
$\beta_{1550}$ & -0.467 & $4.83\times10^{-8}$ & 5.333 & 0.046 & 0.086 & 0.098 \\[3pt]
\rowcolor{gray!40}
$\beta_\mathrm{global}$ & -0.465 & $5.50\times10^{-8}$ & 5.309 & 0.047 & 0.080 & 0.093 \\[3pt]
\rowcolor{gray!30}
$f_\mathrm{esc}^{\mathrm{Ly}\alpha}$ & 0.333 & $1.04\times10^{-4}$ & 3.710 & 0.042 & 0.083 & 0.092 \\[3pt]
\rowcolor{gray!30}
$\tau_1$ & -0.309 & $3.14\times10^{-4}$ & 3.419 & 0.038 & 0.087 & 0.095 \\[3pt]
\rowcolor{gray!30}
$\tau_2$ & -0.298 & $5.09\times10^{-4}$ & 3.286 & 0.046 & 0.080 & 0.092 \\[3pt]
\rowcolor{gray!30}
$f_{\mathrm{esc,~H}\beta}^\mathrm{LyC}$ & 0.293 & $6.68\times10^{-4}$ & 3.208 & 0.041 & 0.088 & 0.097 \\[3pt]
\rowcolor{gray!20}
$\mathrm{sSFR}_{100}$ & -0.275 & 0.001 & 3.005 & 0.057 & 0.097 & 0.113 \\[3pt]
\rowcolor{gray!10}
$\mathrm{sSFR}_{10}$ & -0.207 & 0.015 & 2.158 & 0.057 & 0.087 & 0.104 \\[3pt]
\rowcolor{gray!10}
$\mathrm{EW}(\mathrm{H}\beta)$ & -0.181 & 0.035 & 1.812 & 0.040 & 0.095 & 0.103 \\[3pt]
\rowcolor{gray!10}
$\mathrm{SFR}_{5}$ & -0.177 & 0.039 & 1.761 & 0.051 & 0.087 & 0.101 \\[3pt]
\rowcolor{gray!10}
$\mathrm{SFR}_{10}$ & -0.176 & 0.040 & 1.748 & 0.049 & 0.090 & 0.103 \\[3pt]
\rowcolor{gray!10}
$\mathrm{sSFR}_{5}$ & -0.174 & 0.043 & 1.721 & 0.058 & 0.087 & 0.104 \\[3pt]
$\log_{10} \mathrm{O}_{31}$ & 0.163 & 0.057 & 1.577 & 0.035 & 0.085 & 0.091 \\[3pt]
$\Sigma_{\mathrm{sSFR}_{10}}$ & -0.151 & 0.078 & 1.417 & 0.039 & 0.089 & 0.098 \\[3pt]
$\log \Sigma_{\mathrm{H}\beta}$ & -0.146 & 0.087 & 1.357 & 0.045 & 0.075 & 0.088 \\[3pt]
$\Sigma_{\mathrm{SFR}_{10}}$ & -0.145 & 0.091 & 1.336 & 0.043 & 0.071 & 0.083 \\[3pt]
$\mathrm{SFR}_{100}$ & -0.141 & 0.100 & 1.282 & 0.042 & 0.091 & 0.100 \\[3pt]
$\log Z_{\mathrm{gas}}$ & -0.131 & 0.126 & 1.145 & 0.043 & 0.095 & 0.104 \\[3pt]
$\log U_{\mathrm{gas}}$ & -0.073 & 0.391 & 0.276 & 0.040 & 0.097 & 0.105 \\[3pt]
$f_\mathrm{esc, ~UV}^\mathrm{LyC}$ & 0.072 & 0.404 & 0.244 & 0.057 & 0.087 & 0.104 \\[3pt]
M$_{1500}$ & -0.063 & 0.462 & 0.096 & 0.038 & 0.087 & 0.095 \\[3pt]
$\log_{10} \mathrm{O}_{32}$ & -0.047 & 0.586 & -0.217 & 0.043 & 0.088 & 0.097 \\[3pt]
$\log \Sigma_{1100}$ & -0.032 & 0.706 & -0.543 & 0.040 & 0.081 & 0.090 \\[3pt]
$f_\mathrm{esc, Saldana}^\mathrm{LyC}$ & 0.029 & 0.812 & -0.886 & 0.088 & 0.125 & 0.153 \\[3pt]
$r_{50}$ & 0.023 & 0.790 & -0.806 & 0.053 & 0.094 & 0.108 \\[3pt]
$F_{900}^\mathrm{rest}/F_{1100}^\mathrm{rest}$ & 0.022 & 0.794 & -0.821 & 0.044 & 0.083 & 0.094 \\[3pt]
$\mathrm{EW}(\mathrm{Lya})$ & 0.013 & 0.880 & -1.176 & 0.040 & 0.092 & 0.100 \\[3pt]
\bottomrule
\end{tabularx}
\end{table*}

\subsection{Training Data}
\label{subsec:trainingData}

We generate a training set of synthetic photometric and emission line measurements using \texttt{Prospector}, based on the parameter grid outlined in Table~\ref{tab:sr_training_grid}, resulting in 21870 samples. We decide on generating them based on best performing model \texttt{d2UniNoScale}.

To identify which observables are most informative for predicting the escape fraction, we perform a feature-importance analysis using a Random Forest regressor applied to quantities accessible at high redshift. The resulting feature importances show that the UV $\beta$ slope dominates the predictive power, followed by the Balmer-line diagnostics. Guided by this ranking, we restrict the symbolic regression model to this reduced set of observables, which trace dust attenuation and the ionizing spectrum and are additionally motivated by the parameters exhibiting the strongest Kendall's $\tau$ correlations (Tab.~\ref{tab:tau_results}), together with the widely used [O,\textsc{ii}]/[O,\textsc{iii}] diagnostic. To avoid scale-dependent features, all emission-line fluxes are expressed as dimensionless ratios normalised by H$\beta$; the full feature set is $\{\beta,\,\log_{10}(F_{\rm H\alpha}/F_{\rm H\beta}),\,\log_{10}(F_{[\rm O\,\textsc{iii}]}/F_{[\rm O\,\textsc{ii}]})\}$ together with their squares. We compute weights based on each sample’s location in the prior space of the varied physical parameters, which include stellar metallicity, the relative dust attenuation around young stars ($\tau_1/\tau_2$), diffuse dust attenuation ($\tau_2$), dust attenuation curve slope, gas-phase ionization parameter, and gas-phase metallicity, resulting in a distribution of $-10 < \log_{10}(f_\mathrm{esc}^\mathrm{LyC}) < 0$. The adopted priors are listed in Table~\ref{tab:priors}.

\begin{table}
\centering
\caption{Grid of parameter values used to generate the symbolic regression training set with \texttt{Prospector}.}
\label{tab:sr_training_grid}
\begin{tabular}{ll}
\toprule
\textbf{Parameter} & \textbf{Values} \\
\midrule
$f_\mathrm{scale} $ & 0.0, 0.1, 0.2, 0.3, 0.4, 0.5, 0.6, 0.7, 0.8, 0.9, 1.0 \\
$\log(Z/Z_\odot)$ & $-2.0$, $-1.5$, $-1.0$, $-0.5$, $0.0$ \\
$\log(Z_\mathrm{gas}/Z_\odot)$ & $-2.0$, $-1.5$, $-1.0$, $-0.5$, $0.0$ \\
$\log U_\mathrm{gas}$ & $-3.0$, $-2.5$, $-2.0$, $-1.5$, $-1.0$ \\
$\rm dust~index$ & $-0.7, -0.35, 0.0, 0.2, 0.4$ \\
$\tau_1/\tau_2$ &$ 0.5, 1.0, 1.5$ \\
$\tau_2$ & $0.0, 0.001, 0.01, 0.1, 0.4, 1.0$ \\
\bottomrule
\end{tabular}
\end{table}

\subsection{The Regression Model}
\label{subsec:regressionModel}

During training, we allow the symbolic regression model to use standard binary operators ($+$, $-$, $\times$, $\div$) as well as unary functions including \texttt{exp}, \texttt{log}, and \texttt{abs}. The model is trained by minimizing an element-wise loss function defined as the mean squared error: $L(x, y) = (x - y)^2$. We set a maximum expression size (\texttt{maxsize}) of 20 to balance model complexity and interpretability.

To evaluate model performance, we compare two outputs from the regression process: the overall best model (\textit{best}), which optimally balances accuracy and simplicity, and the most accurate model (\textit{accuracy}), which prioritizes predictive performance.

We experiment with various training set configurations: Specifically, we aim to recover escape fractions as accurately as possible in the observable range, while recognizing that extreme precision is not required for very low values (e.g., distinguishing between $10^{-6}$ and $10^{-20}$). To assess the impact of low escape-fraction values on the symbolic regression training, we consider three strategies: 
(1) excluding data points with $\log_{10}(f_\mathrm{esc}^\mathrm{LyC}) < -3.5$ during training; and
(2) imposing a floor by setting all escape fractions below this threshold to $\log_{10}(f_\mathrm{esc}^\mathrm{LyC}) = -3.5$, retaining their associated physical parameters.
We find that approach (1) most effectively enhances the model’s ability to resolve and distinguish the parameter scales of interest.

\subsection{Regression Results and Linear Regression}

Even when supplied with a broader set of input variables, the simplest expression on the Pareto front that captures the dominant behaviour of $\log_{10}(f_\mathrm{esc}^\mathrm{LyC})$ depends only on $\beta$:

\begin{equation}
    \log_{10}(f_{\rm esc}^\mathrm{LyC}) = a\,\beta + b.
    \label{eq:fesc_beta}
\end{equation}

To obtain accurate coefficients and quantify uncertainties we refine this relation by linear regression via maximum-likelihood estimation (MLE), fitting $a$, $b$, and an intrinsic scatter $\sigma$ simultaneously. Assuming Gaussian intrinsic scatter, the weighted log-likelihood is

\begin{equation}
    \ln\mathcal{L} = -\frac{1}{2}\sum_i w_i
        \left[\frac{\bigl(\log_{10}f_{{\rm esc},i}^\mathrm{LyC} - a\beta_i - b\bigr)^2}{\sigma^2}
        + \ln\!\left(2\pi\sigma^2\right)\right],
\end{equation}

which we maximise with respect to $(a,\,b,\,\sigma)$. Parameter uncertainties are derived from the diagonal of the inverse Hessian evaluated at the maximum, which propagates the degeneracy between scatter and slope into the reported errors. The best-fit relation is

\begin{equation}
    \log_{10}(f_{\rm esc}^\mathrm{LyC}) = (-2.30 \pm 1.28)\,\beta - (6.26 \pm 2.91),
    \quad \sigma = 0.43~\text{dex},
    \label{formula}
\end{equation}

within uncertainties of \cite{chisholm_far-ultraviolet_2022} ($\log_{10}(f_{\rm esc}^{\rm LyC}) = -1.22\,\beta_{\rm obs}^{1550} - 3.89$) The intrinsic scatter of $\sigma = 0.43$~dex reflects the genuine spread in $f_{\rm esc}$ at fixed $\beta$ that a single-observable relation cannot capture. When predicting $f_{\rm esc}$ for an individual galaxy, the total 1$\sigma$ uncertainty on $\log_{10}(f_{\rm esc})$ is obtained by adding the propagated uncertainty on the mean line and $\sigma$ in quadrature.

\begin{figure}
    \centering
    \includegraphics[width=\linewidth]{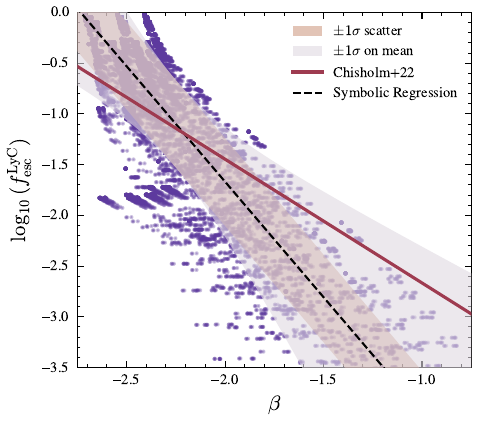}
    \caption{Comparison between $\log_{10}(f_\mathrm{esc}^\mathrm{LyC})$ and the UV continuum slope $\beta$. The black dashed line indicates the relation derived via symbolic regression and subsequently confirmed with a linear regression fit. Dark-violet circles represent the samples generated with \texttt{Prospector}. The orange shaded region illustrates the intrinsic scatter of the relation, while the purple shaded region denotes the uncertainty on the linear regression prediction for the mean trend. The bordeaux solid line shows the $\beta$--$\log_{10}(f_\mathrm{esc}^\mathrm{LyC})$ relation from \citet{chisholm_far-ultraviolet_2022}.}
    \label{fig:placeholder}
\end{figure}

\begin{figure*}
    \centering
    \includegraphics[width=\linewidth]{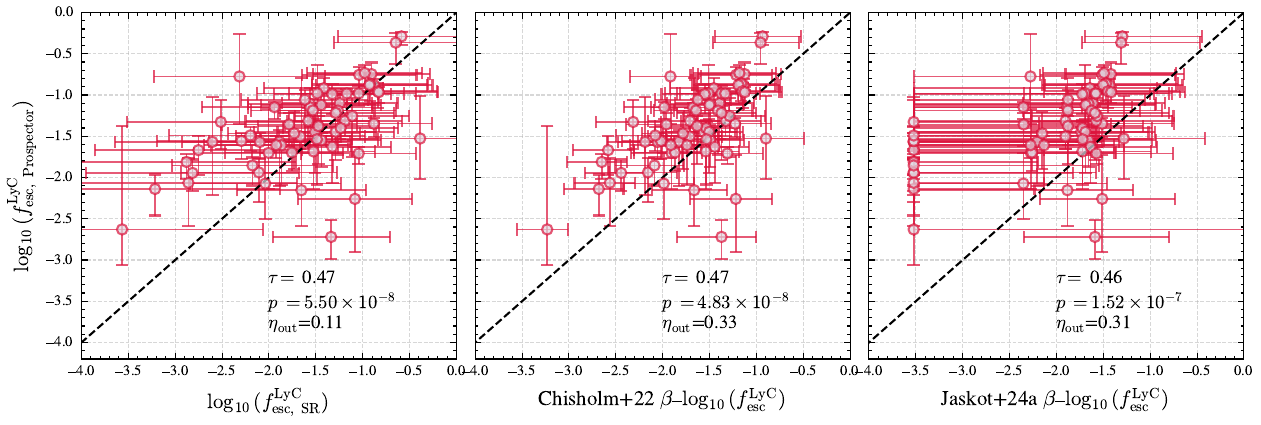}
    \caption{Comparison of the Lyman-continuum escape fraction inferred from the full SED fitting with \texttt{Prospector} (global $f_\mathrm{esc}^\mathrm{LyC}$) to estimates using different methods. The panels show comparisons with (from left to right) to predictions using our $\beta_\mathrm{global}$–$f_\mathrm{esc}^\mathrm{LyC}$ relation determined via SR, predictions based on the $\beta_{1550}$–$f_\mathrm{esc}^\mathrm{LyC}$ relation from \citet{chisholm_far-ultraviolet_2022}, and predictions from the multivariate survival-analysis model of \citet{Jaskot_2024} evaluated using $\beta_{1550}$ alone. The dashed lines indicate one-to-one agreement. Each panel reports Kendall’s $\tau$ correlation coefficient and the associated p-value. All methods yield broadly consistent estimates of $f_\mathrm{esc}^\mathrm{LyC}$ when compared to the Prospector-inferred values, with $\tau$ ranging between 0.46 and 0.47.}
    \label{fig:chisholm}
\end{figure*}

Figure~\ref{fig:chisholm} compares the Lyman-continuum escape fractions inferred from full SED fitting with \texttt{Prospector} to estimates obtained using three $\beta$-based approaches: our symbolic-regression model based on $\beta_\mathrm{global}$, the $\beta_{1550}$–$f_\mathrm{esc}^\mathrm{LyC}$ relation of \citet{chisholm_far-ultraviolet_2022}, and the multivariate survival-analysis model of \citet{Jaskot_2024} evaluated using $\beta_{1550}$ alone.
While the Kendall’s $\tau$ correlation coefficients are nearly identical for all three methods (0.47, 0.47, and 0.46, respectively), the fraction of statistically significant outliers differs substantially. Our regression (Eq.~\ref{formula}) reproduces the \texttt{Prospector}-inferred $f_\mathrm{esc}^\mathrm{LyC}$ within $1\sigma$ for a considerably larger fraction of galaxies, yielding an outlier fraction of 11\%. In contrast, the \citet{chisholm_far-ultraviolet_2022} relation results in an outlier fraction of 38\%, while the survival-analysis relation yields 33\%. The latter is primarily driven by its comparatively larger predictive uncertainties, which broaden the allowed consistency range. Consistent with this trend, the median absolute offset $\Delta \log_{10}(f_\mathrm{esc}^\mathrm{LyC})$ is smallest for the symbolic regression (0.25~dex), intermediate for the \citet{chisholm_far-ultraviolet_2022} relation (0.31~dex), and largest for the \citet{Jaskot_2024} relation (0.61~dex). Together, these results indicate that although all three methods exhibit comparable rank correlations, their quantitative agreement with the \texttt{Prospector} estimates differs significantly, with our regression providing the tightest consistency in both scatter and outlier fraction.
A full comparison between all method and different literature estimates can be found in Fig.~\ref{fig:all_methods} in the Appendix.

Notably, our updated relation between $\beta$ and $f_\mathrm{esc}$ is not valid for UV continuum slopes bluer than $\beta = -2.71$ and should not be extrapolated beyond the range of the training data. It serves as an approximate estimator of $f_\mathrm{esc}$ under specific model assumptions and is intended as a substitute only when full SED fitting is computationally prohibitive. The formula replicates results that can be obtained with greater precision through full SED modelling and should only be applied in contexts where the assumptions about the IMF, isochrones, etc. used in the model -- outlined in Section~\ref{sec:models} -- are appropriate.

\section{Discussion}
\label{sec:discussion}

We have explored a suite of models and priors (Section~\ref{sec:models}), validated their performance with a parameter recovery test (Appendix~\ref{sec:parameterRecoveryTest}), and derived a new set of escape fraction estimates for the LzLCS sample (Table~\ref{tab:escape_fractions}). By comparing the inferred stellar properties to EoR galaxies, enabled by recent JWST observations (Sections~\ref{sec:sfr}--\ref{sec:size}), we confirm that the LzLCS galaxies provide a useful local analogue for high-redshift systems. Importantly, we find that the most extreme LyC leakers do not necessarily exhibit extreme stellar properties.  

To probe potential predictors of $f_\mathrm{esc}^\mathrm{LyC}$, we applied symbolic regression  to high-redshift observables (Section \ref{sec:symreg}). Among the tested relations, the optimal balance of accuracy and interpretability is achieved with a linear relation between $\log_{10}(f_\mathrm{esc}^\mathrm{LyC})$ and the UV continuum slope $\beta$ (Figure~\ref{fig:chisholm}, first panel). While such a correlation has been reported previously \citep{chisholm_far-ultraviolet_2022}, our analysis yields an updated calibration of the relation.

\subsection{Implications}
Our results carry several key implications. 
First, by recovering the observed and unseen $F_\mathrm{LyC}$, we demonstrate that Bayesian inference–based SED fitting can place robust constraints on $f_\mathrm{esc}^\mathrm{LyC}$. The fact that the strongest LyC leakers do not necessarily exhibit the most extreme stellar properties indicates that simple global proxies, such as star formation rate or ionisation state, are insufficient predictors of LyC escape. This underscores the importance of models that capture the complex interplay between stars, gas, and dust. 

Second, the comparison of gas-phase metallicities and stellar properties to both local and high-redshift populations (Sections~\ref{sec:gasPhaseMetal}--\ref{sec:size}) supports the interpretation of the LzLCS galaxies as partial analogues of early galaxies. Although they are offset from local scaling relations, their metallicities and sizes are closer to the properties of systems at $z\sim3$--6. This suggests that the LzLCS provides a valuable bridge population for testing physical models of LyC leakage under conditions intermediate between the nearby Universe and the Epoch of Reionization.  

Finally, the symbolic regression analysis (Section~\ref{sec:symreg}) offers a data-driven way to search for observable predictors of LyC escape. While the updated calibration of the $\beta$–$f_\mathrm{esc}$ relation remains affected by large uncertainties, the inability of the network to identify a simple analytic form involving additional observables (e.g., to break the degeneracy between $\beta$, dust, and nebular properties) suggests that the underlying relations are intrinsically complex and cannot be fully captured by global galaxy properties alone.

\subsection{Caveats}
Although parameters inferred from a common dataset are not statistically independent, the stability of the rank-based correlations under posterior sampling suggests that the observed trends are not dominated by inference-induced degeneracies, while we emphasize that they should not be interpreted as causal.

We explored a range of dust prescriptions to identify the best performing model. However, throughout this work we adopted only a simplified treatment of nebular emission; a parameter that significantly affects $f_\mathrm{esc}^\mathrm{LyC}$.
A key limitation lies in our modelling of nebular emission through pre-computed \texttt{Cloudy} grids, which only account for ionization-bounded regions. This setup assumes that all ionizing photons entering the gas are absorbed and reprocessed, preventing any leakage through low-density, density-bounded channels. As a result, our framework does not currently permit the escape of ionizing photons that would otherwise traverse such channels. Improving the treatment of nebular emission to self-consistently include both ionization- and density-bounded regions remains an important direction for future work.

When comparing the properties of the LzLCS galaxies, we find that many align more closely with the properties of higher-redshift systems than with local analogue galaxies. However, being ``closer'' does not necessarily imply that they are fully representative. One limitation arises from the strong ongoing star formation: young stellar populations can outshine older ones, leaving long tails of possible earlier star formation that cannot be fully excluded. Such hidden older populations may contribute to the integrated properties of a galaxy, yet would be unlikely in systems at $z>6$ due to their age. Furthermore, it remains uncertain whether correlations with $f_\mathrm{esc}^\mathrm{LyC}$ evolve with redshift, or whether galaxies in the Epoch of Reionization are intrinsically different from the local systems available for study.
Future \textit{JWST} observations will enable increasingly spatially resolved studies of LyC escape at $z \gtrsim 2$, where the ionizing continuum is redshifted into the near-UV and optical. At lower redshifts ($z \lesssim 2.3$), however, direct LyC measurements are inaccessible from the ground due to atmospheric absorption, and \textit{JWST} lacks far-UV coverage. In this regime, proposed missions with dedicated far-UV capabilities, such as by the NASA Habitable Worlds Observatory, will be required to directly map LyC escape in nearby galaxies and connect local leakage pathways to global galaxy properties. In the decades leading up to HWO, progress will therefore rely on combining \textit{JWST} IFU observations, indirect LyC tracers, and studies of low-redshift analogs to bridge the gap between resolved local physics and globally inferred escape fractions.

\section{Conclusion}
\label{sec:conclusion}

In this study, we used the Bayesian inference-based SED fitting tool \texttt{Prospector} to analyze the photometric and emission line data of the LzLCS galaxy sample. We explored a suite of physical models with varying assumptions about dust attenuation, star formation history, and the fraction of runaway stars. Among these models, the best-performing configuration for recovering the unobserved LyC flux employed a two-component dust model and a uniform prior between 0 and 1 for the fraction of runaway massive stars ($f_\mathrm{scale}$).

Our results yield a median inferred escape fraction of approximately 4\%, with some galaxies reaching values as high as 51\% and 26 galaxies exhibiting a cosmologically relevant $f_\mathrm{esc}^\mathrm{LyC} \gtrsim5\%$. To test the robustness of our modelling framework, we conducted a parameter recovery test by generating mock observations, synthetic photometry and emission lines with added noise and refitting them using \texttt{Prospector}. We find that the model systematically overpredicts $f_\mathrm{esc}^\mathrm{LyC}$, with the median absolute pull exceeding unity, indicating that the formal posterior uncertainties underestimate the true scatter in the downward direction. We therefore inflate the lower uncertainties by a factor of 1.5, after which the input values are recovered within $1\sigma$. This systematic overprediction is most pronounced at very low escape fractions and dust optical depths, suggesting a shortcoming in the current implementation when neutral hydrogen alone, rather than together with dust, dominates the attenuation of ionizing photons.\\

A likely origin of this limitation lies in the nebular emission modelling used in \texttt{Prospector}, which interpolates from a fixed \texttt{Cloudy}-based grid generated for a specific IMF and ionizing spectrum. The grid assumes ionization-bounded regions only, thereby trapping all ionizing photons within the nebular component and disallowing any escape. This constraint is particularly problematic when modelling LyC leakage through density-bounded channels. 

To provide a computationally inexpensive alternative to full SED fitting and explore correlations between $f_\mathrm{esc}^\mathrm{LyC}$ and other galaxy properties, we trained a symbolic regression model using a grid of synthetic data generated with \texttt{Prospector}. Using only high-redshift-accessible observables — namely the UV $\beta$ slope, the Balmer decrement, and the [O~\textsc{iii}]/[O~\textsc{ii}] ratio — we find that the regression consistently identifies a simple linear relation between $\log_{10}(f_\mathrm{esc})$ and $\beta$ as the optimal solution in terms of complexity and predictive power. A subsequent Bayesian linear fit for the scatter yields $\log_{10}(f_\mathrm{esc}) = (-2.3 \pm 1.28) \cdot \beta - (6.26 \pm 2.91), \sigma = 0.43~$dex. This relation should only be applied within the validity range of our model assumptions (see Sec.~\ref{sec:models}) and must not be extrapolated beyond $\beta < -2.71$, where symbolic regression suggests a diversion from a linear relation.\\

Overall, our results highlight both the potential and the limitations of inferring LyC escape fractions from indirect observables. While ratios like $f_\mathrm{esc}$ were originally conceived to eliminate certain observational uncertainties, the anisotropic nature of LyC escape, which is well established through simulations and lensing-based observations, raises concerns about interpreting such quantities from a single line of sight.

\section*{Acknowledgments}

We thank William McClymont for insightful discussion. We thank Sophia Flury for making her data available. AS thanks the Science and Technology Facilities Council (STFC) Center for Doctoral Training (CDT) in Data Intensive Science at the University of Cambridge (STFC grant number ST/W006812/1) for a PhD studentship which is also partly funded by the UK Research and Innovation (UKRI) Frontier Research grant RISEandFALL. Various software packages were used in this work, including
numpy \citep{2020NumPy-Array}, scipy \citep{2020SciPy-NMeth}, matplotlib \citep{Hunter:2007}, and astropy \citep{2013A&A...558A..33A, 2018AJ....156..123A, 2022ApJ...935..167A}. Language editing and figure formatting were supported by ChatGPT \citep{openai_chatgpt}.

\section*{Data Availability}

The data underlying this article, including the derived SED-fitting products, are publicly available via the \texttt{LzLCS\_Prospector} GitHub repository at \url{https://github.com/Espe13/LzLCS_Prospector}.



\bibliographystyle{mnras}
\bibliography{example} 




\appendix

\onecolumn

\section{Parameter Recovery Test}
\label{sec:parameterRecoveryTest}

To evaluate the predictive performance of \texttt{PROSPECTOR}, we conducted a three-step parameter recovery test (PRT), based on the \texttt{d2uniNoScale} model described in Sect.~\ref{sec:models}, which showed the highest accuracy in recovering the observations.

\subsection{Mock Data}
First, we generated mock photometry and spectroscopy using an input parameter set ($\Vec{\theta}_\text{in}$) with \texttt{PROSPECTOR}. We then added Gaussian noise to this data to simulate realistic observational uncertainties. Finally, we fit the noisy data to recover the output parameter vector ($\Vec{\theta}_\text{out}$), comparing it to $\Vec{\theta}_\text{in}$ to evaluate prediction accuracy.\\
Mock photometry included SDSS \textit{ugriz} and GALEX \textit{FUV} and \textit{NUV} bands, while emission lines were selected from Flury’s study \cite{flury_low-redshift_2022}, as described in Section~\ref{subsec:photEL}. We applied Gaussian noise to the line intensities, using an H$\beta$ signal-to-noise ratio (SNR) to scale the noise for all lines:\\
\begin{align}
\Phi_\text{mock} = \Phi_\text{predicted} + \mathcal{N}(0, \sigma = \text{H}_\beta/\text{SNR})
\end{align}

For spectroscopic data, we enforced a minimum uncertainty of 5\% of the emission line flux during fitting. Photometric noise was added with fixed SNRs: 20 for SDSS and 10 for GALEX. All emission line SNR references pertain to H$\beta$, with photometric SNRs held constant.\\
We performed 11 PRTs varying frac\_obrun from 0 to 1 in steps of 0.1 (resulting in a variation of $f_\mathrm{esc}^\mathrm{LyC}$ between 0 and 35\%), across SNR values of 1, 3, 5, 10, and 30. Each combination was fit ten times, totaling 550 tests, to ensure statistical robustness. To adapt the test to the LzLCS, we assumed $z = 0.34$.

\subsection{Mock Results}
\label{subsec:mockResult}

\begin{figure}
    \centering
    \includegraphics[width=0.5\linewidth, alt={The median predicted escape fractions follow the 1:1 line with a slight overestimation bias, indicating a tendency to recover higher escape fractions than the true values.}]{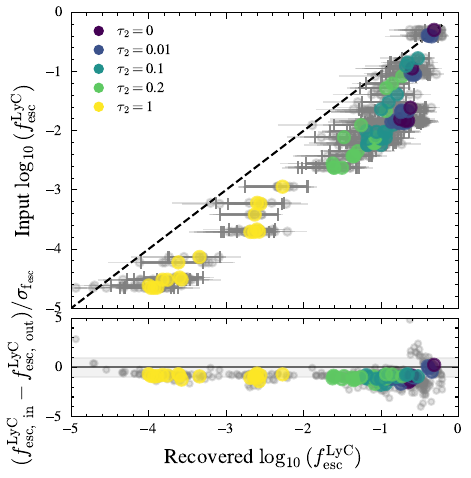}
    \caption{\textbf{Recovery of the input Lyman continuum escape fraction from mock tests using \texttt{Prospector}.}
    The upper panel compares the predicted versus true escape fractions (both in logarithmic scale), with the dashed line indicating perfect recovery. Individual mock realizations are shown as light gray points; colored symbols denote the median per input $f_\mathrm{esc}^\mathrm{LyC}$ bin, with the color encoding the input stellar dust optical depth $\tau_2$. The lower panel shows the pull, $(f_\mathrm{esc,in}^\mathrm{LyC} - f_\mathrm{esc,~out}^\mathrm{LyC})/\sigma_{f_\mathrm{esc}}$, with the shaded band marking $|\mathrm{pull}| \leq 1$. Considering all test cases, the predicted escape fraction is overestimated by $0.8\,\mathrm{dex}$ on average; restricting to $f_\mathrm{esc,in} \geq 10^{-3}$ reduces this bias to $0.6\,\mathrm{dex}$. The median pull of $-1.45$ indicates a mild systematic underestimate of the reported uncertainties, yet the overall recovery remains adequate for the purposes of this work.}
    \label{fig:escape_recovery}
\end{figure}

Figure \ref{fig:escape_recovery} compares the input $f_\mathrm{esc}^\mathrm{LyC}$ with the one we recover after fitting the mockservations with the model described in Section~\ref{subsec:log20} for different dust input values. The predicted escape fraction is overestimated by $0.8\,\mathrm{dex}$ on average; restricting to $f_\mathrm{esc,in} \geq 10^{-3}$ reduces this bias to $0.6\,\mathrm{dex}$.
The model particularly is challenged by low escape fractions in combination with very little dust attenuation.

\begin{figure*}
    \centering
    \includegraphics[width=\linewidth]{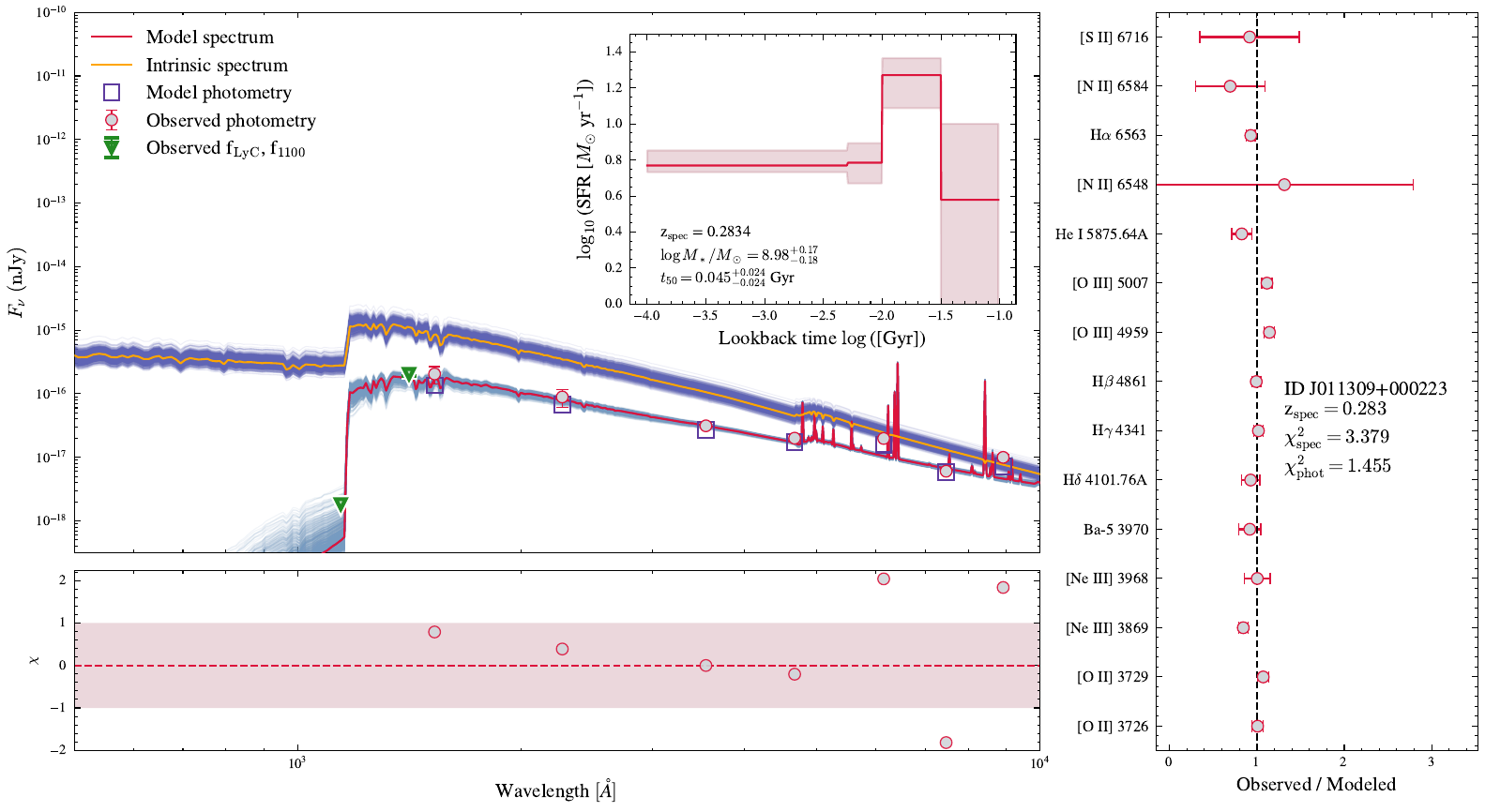}
    \caption{Left: best-fit model spectrum (red) with its uncertainty (light blue), alongside observed photometry (white circles, red edge), UV flux measurements (green triangles), and synthetic photometry (purple squares). The intrinsic, unattenuated spectrum is shown in orange with associated uncertainties (purple). The inset shows the corresponding star formation history, with lookback time in $\log$~Gyr on the horizontal axis and $\log_{10}(\mathrm{SFR}\,[M_\odot\,\mathrm{yr}^{-1}])$ on the vertical axis. Right: emission-line fit for the same object, expressed as the ratio of observed to predicted flux for each fitted emission line.}
    \label{fig:fit_combined}
\end{figure*}

\begin{figure*}
    \centering
    \includegraphics[width=\linewidth]{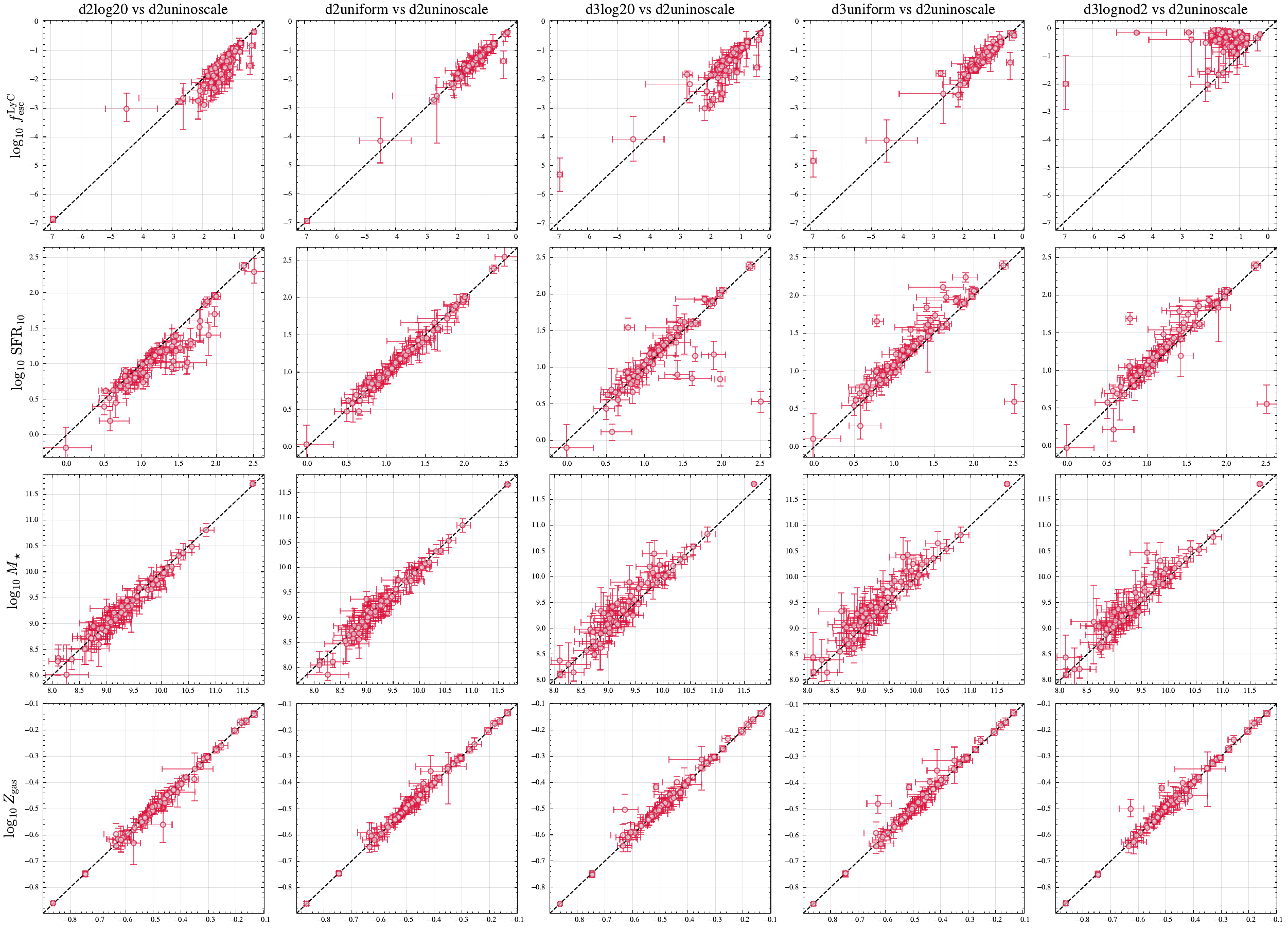}
    \caption{Comparison of parameter estimates from our best-performing model (\texttt{d2uniNoScale}) with those obtained using alternative models (\texttt{d2uni}, \texttt{d2log20}, \texttt{d3log20}, \texttt{d3uniform}, and \texttt{d2log20nod2}). Shown are results for stellar mass, gas-phase metallicity, SFR$_{10}$, and $\log(f_\mathrm{esc}^\mathrm{LyC})$. All models yield consistent estimates within $2\sigma$ for all parameters, with the exception of the picket-fence–like \texttt{d2log20nod2}, which systematically predicts higher escape fractions.}
    \label{fig:compare_all}
\end{figure*}

\begin{figure*}
    \centering
    \includegraphics[width=0.9\linewidth]{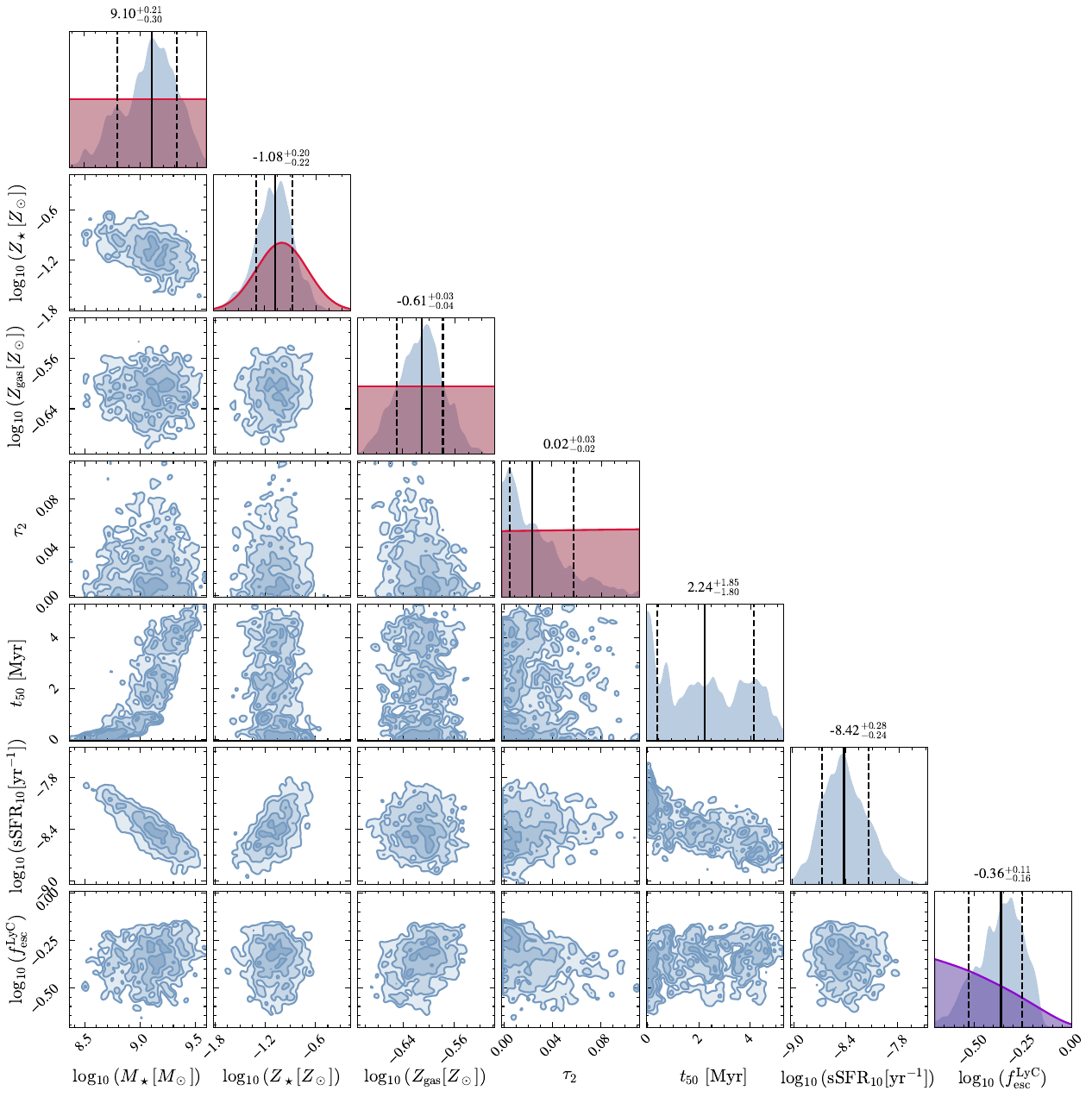}
    \caption{Posterior distributions (blue) for the best-fit model of galaxy J011309+000223 by the model \texttt{d2uniNoScale}, which has a high $f_\mathrm{esc}^\mathrm{LyC} = 51^{+6}_{-11}\%$. The diagonal panels show marginalized one-dimensional distributions, with vertical lines indicating the median and the 16th and 84th percentiles. Off-diagonal panels display the two-dimensional parameter correlations. We present directly fitted parameters (stellar mass, stellar metallicity, gas-phase metallicity and optical depth of the diffuse dust component ($\tau_2$)) alongside their priors (orange), as well as derived quantities: the half-mass assembly time ($t_{50}$), the specific SFR over the past 10 Myr ($\log_{10}(\mathrm{sSFR}_{10})$), and $f_\mathrm{esc}^\mathrm{LyC}$ with its derived prior. The comparison between posteriors and priors demonstrates that the fitted parameters are informed by the data and not solely driven by the assumed priors. We find the expected degeneracies between stellar mass, metallicity and dust.}
    \label{fig:corner_highfesc}
\end{figure*}

\begin{figure*}
    \centering
    \includegraphics[width=\linewidth]{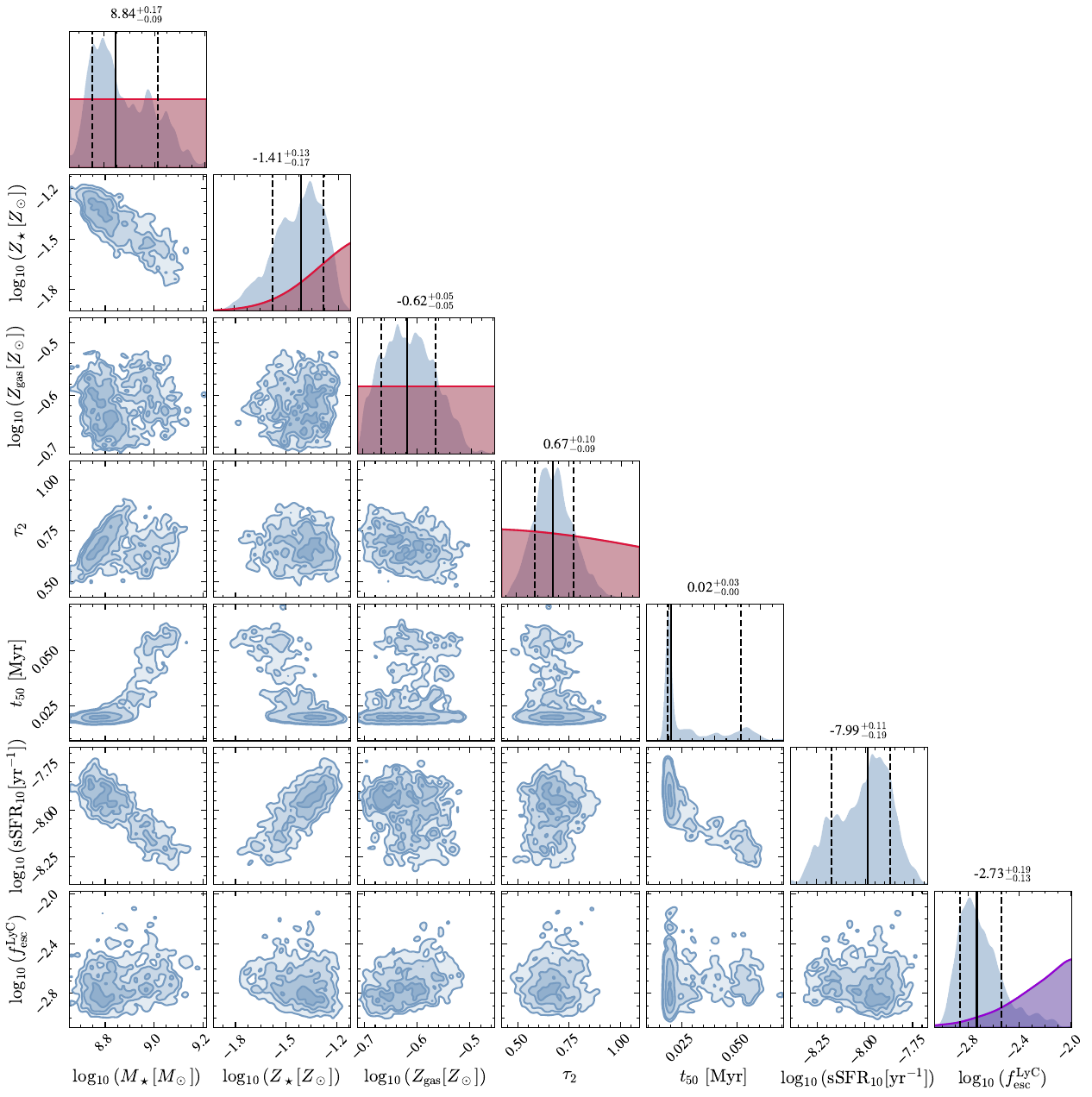}
    \caption{Same as Fig.~\ref{fig:corner_highfesc}, for galaxy J124033+212716, which has a low $f_\mathrm{esc}^\mathrm{LyC} = 0.1^{+0.0}_{-0.1}\%$. }
    \label{fig:corner}
\end{figure*}

\subsection{Symbolic Regression Result Comparison}
\label{subsec:symbolicregression}

\begin{figure}
    \centering
    \includegraphics[width=0.9\linewidth]{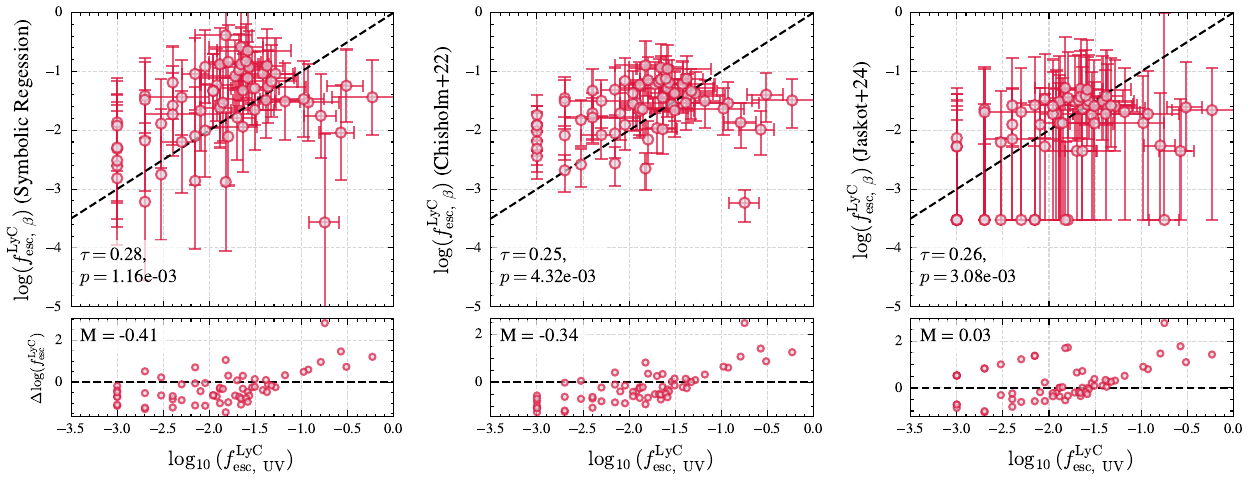}
    \vspace{0.3cm}
    \includegraphics[width=0.9\linewidth]{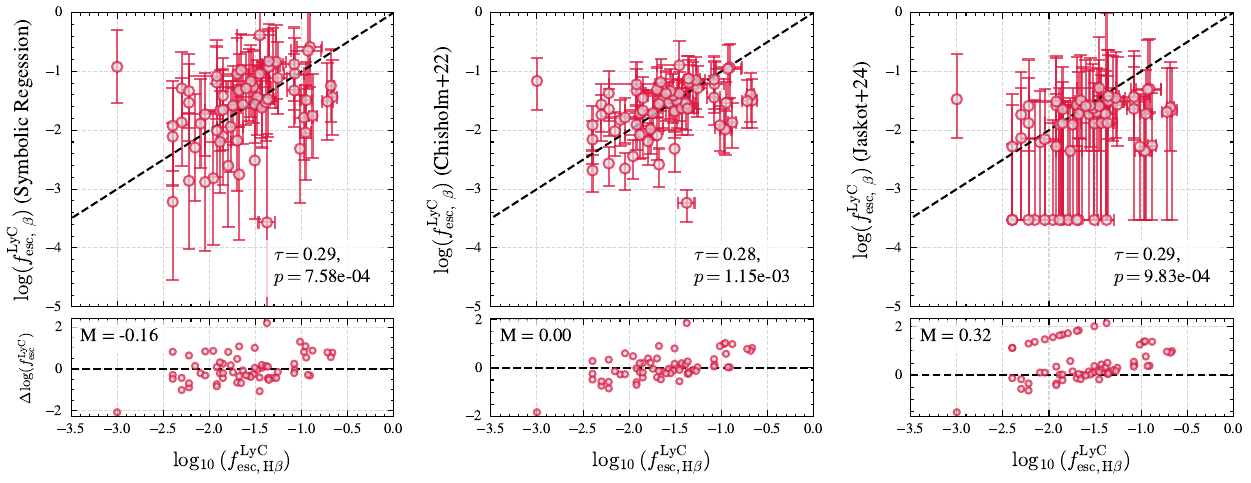}
    \vspace{0.3cm}
    \includegraphics[width=0.9\linewidth]{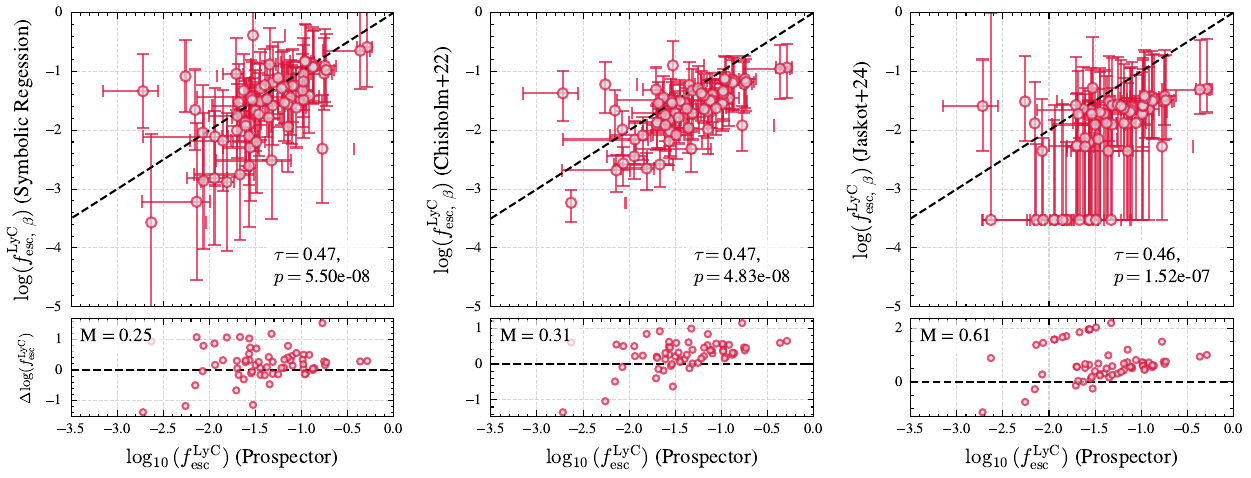}
    \caption{
    Comparison of $\log_{10}(f_{\mathrm{esc}}^{\mathrm{LyC}})$ inferred from H$\beta$-based estimates, $\log_{10}(f_{\mathrm{esc,~H}\beta}^{\mathrm{LyC}})$ (top row), UV based estimations $\log_{10}(f_{\mathrm{esc,~UV}}^{\mathrm{LyC}})$ (second row), and \texttt{Prospector}-based estimations $\log_{10}(f_{\mathrm{esc,~Prospector}}^{\mathrm{LyC}})$ (third row), with three $\beta$-based predictions: our symbolic-regression model (left), the $\beta$–$f_{\mathrm{esc}}$ relation of \citet{chisholm_far-ultraviolet_2022} (middle), and the $\beta$-based prediction from the survival-analysis model of \citet{Jaskot_2024} evaluated using $\beta$ alone (right). Points show individual galaxies with asymmetric uncertainties propagated in log space; the dashed line indicates one-to-one agreement. The lower panels show residuals, $\Delta \log_{10}(f_{\mathrm{esc}}^{\mathrm{LyC}})=\log_{10}(f_{\mathrm{esc}}^{\mathrm{LyC}})-\log_{10}(f_{\mathrm{esc},\beta}^{\mathrm{LyC}})$, with the median residual $M$ reported in each panel. Each top panel lists Kendall’s $\tau$ correlation coefficient and the corresponding $p$-value.
    }  
    \label{fig:all_methods}
\end{figure}

\bsp	
\label{lastpage}
\end{document}